\pdfoutput=1

\documentclass[11pt,a4paper]{article}
\usepackage[normalem]{ulem}

\usepackage{jheppub}
\usepackage{amsmath}
\usepackage{amssymb}
\usepackage{appendix}
\usepackage{multirow}
\usepackage{longtable} 
\usepackage{booktabs} 
\usepackage{adjustbox}
\usepackage{rotating}
\usepackage{diagbox}
\usepackage{float}
\usepackage{xcolor}
\usepackage{cancel}
\usepackage{tabularx}
\usepackage[capitalize]{cleveref}
\usepackage{soul}
\usepackage{url}
\usepackage{hyperref}
\usepackage{colortbl}

\usepackage{color}

\newcommand{\bea}{\begin{eqnarray}}
\newcommand{\eea}{\end{eqnarray}}

\newcommand{\Cc}[1]{{\mathcal C}_{#1}}

\usepackage{color}
\definecolor{niceblue}{rgb}{0,0,1}
\definecolor{nicered}{rgb}{0.7,0.1,0.1}
\definecolor{nicegreen}{rgb}{0.1,0.5,.1}

\usepackage{graphicx}
\usepackage{subcaption}
\graphicspath{{figures/}{./}}

\title{To (b)e or not to (b)e: No electrons at LHCb}

\author[a]{Marcel Alguer\'o,}
\author[a]{Aritra Biswas,}
\author[a,b]{Bernat Capdevila,}
\author[c]{S\'ebastien Descotes-Genon,}
\author[a]{Joaquim Matias,}
\author[d]{Martin Novoa-Brunet}

\affiliation[a]{Universitat Aut\`onoma de Barcelona, 08193 Bellaterra, Barcelona,\\
Institut de F\'{i}sica d'Altes Energies (IFAE), The Barcelona Institute of Science and Technology, Campus UAB, 08193 Bellaterra (Barcelona)}
\affiliation[b]{DAMTP, University of Cambridge, Wilberforce Road, Cambridge, CB3 0WA, United Kingdom}
\affiliation[c]{Universit\'e Paris-Saclay, CNRS/IN2P3, IJCLab, 91405 Orsay, France}
\affiliation[d]{Istituto Nazionale di Fisica Nucleare, Sezione di Bari, Via Orabona 4, I-70126 Bari, Italy}

\abstract{We discuss the impact of the recent LHCb update on the two lepton-flavour universality ratios $R_K$ and $R_{K^*}$, and the CMS update of $B({B_s \to \mu^+\mu^-})$ regarding the possibility of New Physics in $b\to s\ell^+\ell^-$ decays. We perform global fits of the New Physics Wilson coefficients defined in the model-independent approach of the Weak Effective Theory at the $b$-quark mass. We discuss three different frameworks for this analysis: i) an update limited to the experimental data using the same theoretical framework as in earlier works, ii) a full update concerning both the experimental inputs and the theoretical framework, iii) an analysis without the LHCb results on electron modes. The comparison between these sets of results allows us to identify the differences stemming from the various components of the analysis: new experimental results, new inputs for the hadronic form factors, the role played by LHCb data on electron modes. As expected, the significance of all New Physics hypotheses gets reduced after the LHCb announcements on $R_{K^{(\ast)}}$ while the hypothesis of a lepton-flavour-universal contribution to the Wilson coefficient of the semileptonic $O_{9\ell}$ operators (possibly with a very small lepton-flavour-universality violating component) is reinforced. We also discuss the possibility of a long-distance charm-loop contribution through a mode-by-mode analysis and we find that the preferred values for the $\Cc{9\mu}$ Wilson coefficient are consistent throughout the different $b\to s\mu^+\mu^-$ modes and that there is no significant evidence of non-constant $q^2$ dependencies, which would indicate the presence of a long-distance charm-loop contribution beyond those already included.

}

\emailAdd{malguero@ifae.es}
\emailAdd{abiswas@ifae.es}
\emailAdd{bernat.capdevila.soler@gmail.com}
\emailAdd{sebastien.descotes-genon@ijclab.in2p3.fr}
\emailAdd{matias@ifae.es}
\emailAdd{martin.novoa@ba.infn.it}

\begin{document}

\begin{flushright} {\small\scshape BARI-TH/23-747}\end{flushright}
\vspace{-28pt}
\maketitle
\section{Introduction}
\label{intro}

The Flavour-Changing Neutral-Current (FCNC) $b\to s\ell^+\ell^-$ decays have been under intense scrutiny for over more than a decade in relation with their potential sensitivity to New Physics (NP)~\cite{Albrecht:2021tul,London:2021lfn}. Deviations have been observed in $b\to s\mu^+\mu^-$ decays in a consistent way by the LHCb collaboration, not only in branching ratios for $B\to K\mu^+\mu^-$, $B\to K^*\mu^+\mu^-$ and $B_s\to\phi\mu^+\mu^-$~\cite{LHCb:2014cxe,LHCb:2016ykl,LHCb:2021zwz}, but also in $B\to K^*\mu^+\mu^-$ angular observables~\cite{LHCb:2020lmf}. Measurements were also performed by the ATLAS, CMS and Belle collaborations for the same modes~\cite{CMS:2013mkz,CMS:2015bcy,CMS:2017rzx,ATLAS:2018gqc,CMS:2020oqb}.
The Standard Model (SM) prediction of these observables is certainly difficult due to hadronic uncertainties, even though it has benefited from progress in the determination of form factors~\cite{Ball:2004rg,Horgan:2013hoa,Bouchard:2013eph,Bharucha:2015bzk,Gubernari:2018wyi} as well as non-local charm-loop contributions~\cite{Khodjamirian:2010vf,Khodjamirian:2012rm,Gubernari:2020eft}. This has led to the definition of optimised angular observables~\cite{Matias:2012xw,Descotes-Genon:2013vna,Descotes-Genon:2013wba}, with a more limited sensitivity to hadronic effects, allowing for a cleaner interpretation of deviations in terms of NP, in particular in the so-called $P_5'$ observable~\cite{Descotes-Genon:2012isb} exhibiting persistent deviations with respect to SM expectations. Other modes and observables have also been analysed~\cite{LHCb:2018jna,CMS:2018qih,LHCb:2020gog,LHCb:2021xxq}, in particular $B_s\to\mu^+\mu^-$~\cite{LHCb:2021vsc,LHCb:2021awg,ATLAS:2018cur,CMS:2022mgd}
 as well as modes related to the companion radiative decay $b\to s\gamma$~\cite{LHCb:2013pra,LHCb:2020dof,LHCb:2021byf}.

Another type of observables has been the subject of much attention, namely ratios (or differences) of $b\to s\ell^+\ell^-$ observables probing Lepton Flavour Universality (LFU) between electron and muon modes~\cite{Hiller:2003js,
Capdevila:2016ivx,
Bifani:2018zmi}. A very encouraging observation from the LHCb collaboration was the hint of deviations in the LFU ratios $R_K$ and $R_{K^*}$ at low dilepton invariant mass~\cite{LHCb:2014vgu,LHCb:2017avl,LHCb:2021trn}, since these ratios are predicted very accurately within the SM, even after including QED radiative effects~\cite{Isidori:2022bzw}. Similar LFU observables for branching ratios~\cite{BELLE:2019xld,Belle:2019oag} and angular observables~\cite{Belle:2016fev} were also measured at Belle, as well as other modes at LHCb~\cite{LHCb:2019efc,LHCb:2021lvy}, but the limited accuracy of these measurements prevents one from drawing any clear-cut conclusion.

All these measurements can be analysed in a model-independent way, using the Weak Effective Theory (WET) obtained by splitting the dynamics of the SM and its extensions between low and high energies~\cite{Grinstein:1987vj,Buchalla:1995vs}. Short-distance degrees of freedom are integrated out to yield numerical Wilson coefficients, whereas long-distance ones remain dynamical and are involved in local operators. NP scenarios can be expressed through contributions to the Wilson coefficients (which can vanish in the SM), and global fits can be used to assess which scenarios are able to describe the pattern of deviations (and non-deviations) observed. Over the recent years, the accumulation of data on $b\to s\mu^+\mu^-$ observables has progressively led to scenarios involving NP contributions to vector/axial operators for $b\to s\mu^+\mu^-$ (i.e. ${\cal C}_{9\mu}$ and other Wilson coefficients like ${\cal C}_{10\mu}$ and ${\cal C}_{9'\mu}$). Scenarios involving also  universal contributions  to $b\to se^+e^-$ and $b\to s\mu^+\mu^-$, which were
first proposed in Ref.~\cite{Alguero:2018nvb}, provided an even better description of the data (for a recent analysis and further references, see Ref.~\cite{Alguero:2021anc}). Interestingly, the groups working on these global fits reached very similar conclusions at that time, in spite of differences in the choice of theoretical inputs and statistical treatments~\cite{Altmannshofer:2021qrr,Geng:2021nhg,Hurth:2021nsi,Alok:2019ufo,Ciuchini:2020gvn,Datta:2019zca}.

Recently, the LHCb collaboration has announced new results for the two ratios $R_K$ and $R_{K^*}$ each in two bins of the dilepton invariant mass, based on 9 fb$^{-1}$ of data~\cite{LHCb:2022zom,LHCb:2022qnv}. All four measurements show no deviations from the SM expectations, in stark contrast with the earlier LHCb measurements. LHCb can identify more easily the properties of muons than electrons, so that the observables involving electrons may be affected by large experimental systematics (related to missing backgrounds and particle identification) that have proved difficult to estimate and have required a long dedicated experimental effort to pin down. The phenomenological implications of this result w.r.t selected WET, LFU and LFU Violating (LFUV) scenarios as well as within the framework of specific dynamical models has been studied in ref.~\cite{Greljo:2022jac}.
These new results have an impact on the NP scenarios assuming a large LFUV between muons and electrons, as the new data suggest NP contributions of similar sizes to both types of modes. However, one should stress that these results do not affect the picture of deviations in $b\to s\mu^+\mu^-$ observables, which still remains to be explained even if the previous hints of LFUV are far less significant now. It is thus important not only to assess the impact of these new results on the global fits performed up to now for NP in $b\to se^+e^-$ and $b\to s\mu^+\mu^-$, but also to reconsider the impact of the $b\to s\mu^+\mu^-$ observables alone, which should not be affected by the large experimental systematics for electrons that are challenging to determine.

\newpage

In this article we will perform several series of fits, introducing changes on both the experimental and theoretical sides step by step, so that the reader can gauge the impact of each change by comparing the different sets of fits.
In Sec.~\ref{sec:LHCb_global}, we perform a first update of Ref.~\cite{Alguero:2021anc} by including the recent results on $R_K$, $R_{K^*}$ and $B_s\to\mu^+\mu^-$, but keeping the same theoretical inputs. We discuss the modifications induced by this partial update of our framework in a schematical way.
In Sec.~\ref{sec:fullupdate}, we fully update our analysis by considering not only the experimental measurements included in the previous section, but also the theoretical improvements  concerning the determination of form factors from $B$-meson Light-Cone Sum Rules~\cite{Gubernari:2018wyi,Gubernari:2020eft} and lattice computations~\cite{Parrott:2022rgu,Parrott:2022dnu}. The results obtained in this fully updated framework should be considered as the main outcome of this article. 
In Sec.~\ref{sec:electronless}, 
given the systematics attached to electrons at LHCb, we 
change the perspective and 
explore global fits which do not involve LHCb measurements with electrons. At the end of this section, 
 we predict the LFU ratios $R_X$ (instead of taking them as inputs) for a specific LFU NP scenario which is favoured by current data and can be connected with the deviations observed in $R_{D^{(\ast)}}$ 
 ratios.
In Sec.~\ref{charm} we perform a mode-by-mode and bin-by-bin extraction of ${\cal C}_{9\mu}$ to check that all values are consistent with each other and do not exhibit a dependence on the mode or the bin, in opposition to the expected behaviour of a long-distance charm-loop contribution.
 We draw a few conclusions in Sec.~\ref{sec:conclusions}. 
 The appendices provide further detail on:
i) 
the results of our updated global fits without LHCb data on electron-dependent observables from LHCb  in App.~\ref{app:noElectrons} and
 ii) a summary of our updated SM predictions with our updated experimental inputs and theoretical framework in App.~\ref{app:SMpredictions}.

\section{Partial update including recent experiment results }
\label{sec:LHCb_global}

In this section we present partially updated results of the global fits for the NP scenarios already studied in our previous articles~\cite{Descotes-Genon:2015uva,Capdevila:2017bsm,Alguero:2019ptt,Alguero:2021anc}. At this stage, we include the experimental inputs made available recently, while keeping the same theoretical inputs for the form factors as well as for the non-local charm-loop contribution that were used in Ref.~\cite{Alguero:2021anc}.
We provide this series of fits for the purpose of comparison with our earlier work, so that the impact of the updated experimental inputs can be isolated. We stress however that the results discussed in this section should not be considered as the final outcome of our analysis, which will be given in Sec.~\ref{sec:LHCb_global_results}. 

\subsection{Experimental updates}

We thus consider the same experimental inputs as in Ref.~\cite{Alguero:2021anc}, up to the following updates:
\begin{itemize}
\item We consider the four new LHCb inputs for $R_K$ and $R_{K^*}$ given in Refs.~\cite{LHCb:2022zom,LHCb:2022qnv}, including their correlations. Accordingly, we remove the previous LHCb measurements for these quantities~\cite{LHCb:2017avl,LHCb:2021trn}. 
\item We include the most recent CMS result on $B({B_s \to \mu^+\mu^-})$~\cite{CMS:2022mgd} and we follow Ref.~\cite{Hurth:2022lnw} for the average with the determinations from LHCb and ATLAS~\cite{LHCb:2021vsc,LHCb:2021awg,ATLAS:2018cur} leading to the experimental average $B({B_s \to \mu^+\mu^-})=(3.52^{+0.32}_{-0.30})\times 10^{-9}$.
\item In the particular case of scenarios relating neutral ($b\to s\ell^+\ell^-$)  and charged $(b\to c\ell\nu$) anomalies, we consider the most recent HFLAV average for $R_{D^{(*)}}$ (see Ref.~\cite{HeavyFlavorAveragingGroup:2022wzx} and the 2023 online update): $R_D = 0.356 \pm 0.029$ and $R_{D^*} = 0.284 \pm 0.013$ with a correlation coefficient $\rho = -0.37$, which includes both the combined measurement performed in 2022~\cite{LHCb:2023zxo} and the 2023 measurement of $R_{D^*}$, with hadronic $\tau^+$ decays $\tau^+\to \pi^+\pi^-\pi^+(\pi^0)\bar{\nu}_\tau$, by the LHCb collaboration~\cite{LHCb:2023cjr}.
\end{itemize}
All previous observables are also included, together with the electronic modes. In particular, as LHCb has not performed the analysis of the whole 9 fb$^{-1}$ data set for the individual $B\to K^{(*)}\ell^+\ell^-$ branching ratios (only LFU ratios are available currently with the full data set), we keep our earlier inputs for the muonic branching ratios.

\subsection{Fit results with the experimental update}

We consider the same 1D, 2D, 6D and LFU fits as in Ref.~\cite{Alguero:2021anc}. All these scenarios can be expressed in terms of the Wilson coefficients at the scale $\mu_b=4.8$ GeV, where we separate the SM and NP contributions as $\Cc{i\ell} = \Cc{i\ell}^{\rm SM} + \Cc{i\ell}^{\rm NP}$, and we split in some hypotheses the NP contribution further in an LFU (Universal) and an LFUV (Violating) part according to $\Cc{ie}=\Cc{i}^{\rm U}$ and $\Cc{i\mu}=\Cc{i}^{\rm U}+\Cc{i\mu}^{\rm V}$.

For all 1D and 2D scenarios (see \cref{tab:results1D_Marcel,tab:results2D_Marcel}), one can immediately notice a reduction of Pull$_{\rm SM}$ w.r.t. Ref.~\cite{Alguero:2021anc}. This is expected as the new values of $R_K$ and $R_{K^*}$ from LHCb are compatible with the SM at less than $1\sigma$. This is further reflected by the Pull$_{\rm SM}$ values ($1.5\sigma$ or less) obtained for the ``LFUV'' fits for the 1D and 2D scenarios. 

\begin{table}[H] 
    \centering
    \begin{adjustbox}{width=1.\textwidth,center=\textwidth}
\begin{tabular}{c||c|c|c|c||c|c|c|c} 
 & \multicolumn{4}{c||}{All} &  \multicolumn{4}{c}{LFUV}\\
\hline
1D Hyp.   & Best fit& 1$\sigma$/2$\sigma$   & Pull$_{\rm SM}$  & p-value & Best fit & 1$\sigma$/ 2$\sigma$  & Pull$_{\rm SM}$ & p-value\\
\hline\hline
\multirow{2}{*}{$\Cc{9\mu}^{\rm NP}$}    & \multirow{2}{*}{-0.66} &    $[-0.82,-0.50]$ &    \multirow{2}{*}{4.2}   & \multirow{2}{*}{23.9\,\%}
&   \multirow{2}{*}{-0.22}   &$[-0.41,-0.05]$&   \multirow{2}{*}{1.3}  & \multirow{2}{*}{90.5\,\%}  \\
 &  & $[-0.97,-0.34]$ &  & &  &  $[-0.60,+0.12]$ & \\
 \multirow{2}{*}{$\Cc{9\mu}^{\rm NP}=-\Cc{10\mu}^{\rm NP}$}    &   \multirow{2}{*}{-0.19} &    $[-0.25,-0.13]$ &   \multirow{2}{*}{3.0}  & \multirow{2}{*}{14.1\,\%}
 &  \multirow{2}{*}{-0.09}   &   $[-0.16,-0.02]$ & \multirow{2}{*}{1.2}   & \multirow{2}{*}{90.1\,\%}  \\
 &  & $[-0.32,-0.06]$ &  & & & $[-0.23,+0.05]$  &    \\
 \multirow{2}{*}{$\Cc{9\mu}^{\rm NP}=-\Cc{9'\mu}$}     & \multirow{2}{*}{-0.67} &    $[-0.87,-0.47]$   &  \multirow{2}{*}{3.7}  & \multirow{2}{*}{19.0\,\%}
 &  \multirow{2}{*}{-0.05}   &    $[-0.31,+0.16]$  & \multirow{2}{*}{0.2} & \multirow{2}{*}{84.6\,\%} \\
 &  & $[-1.22,-0.59]$ &  & & & $[-0.64,+0.34]$ &    \\
\end{tabular} \end{adjustbox}
\caption{Experimental update only. Most prominent 1D patterns of NP in $b\to s\mu^+\mu^-$. Pull$_{\rm SM}$ is quoted in units of standard deviation. The p-value of the SM hypothesis is $8.0\%$ for the fit ``All'' and $87.6\%$ for the fit LFUV.} 
\label{tab:results1D_Marcel}
\end{table}

In our previous analyses (see for instance Ref.~\cite{Alguero:2021anc}), the two 1D scenarios $\Cc{9\mu}^{\rm NP}$ and $\Cc{9\mu}^{\rm NP} = -\Cc{10\mu}^{\rm NP}$ exhibited rather close Pull$_{\rm SM}$, with a small preference for $\Cc{9\mu}^{\rm NP}$ in the ``All'' fit and the other way around for the ``LFUV'' fit. In Table~\ref{tab:results1D_Marcel}, the preference for a scenario $\Cc{9\mu}^{\rm NP}$ is now clearer for the ``All'' fit. The two scenarios present almost the same (very small) Pull$_{\rm SM}$ for the ``LFUV'' fit. This is mainly driven by the current SM-like value of $B({B_s \to\mu^+\mu^-})$. Even the scenario $\Cc{9\mu}^{\rm NP}=-\Cc{9'\mu}$ fits the current set of global data slightly better than $\Cc{9\mu}^{\rm NP} = -\Cc{10\mu}^{\rm NP}$, due to the SM-like value for $R_K$ in this scenario. With the current LHCb data, the $2\sigma$ intervals of the ``LFUV'' fits for all these scenarios are now compatible with the SM.

In the case of the 2D fits (see Table~\ref{tab:results2D_Marcel}), we notice a global reduction of the Pull$_{\rm SM}$ of around 2.8$\sigma$ for all scenarios, apart from two cases exhibiting a larger pull than expected compared to this global downward trend, namely Hypothesis 2 ($\Cc{9\mu}^{\rm NP}=-\Cc{9^\prime\mu} , \Cc{10\mu}^{\rm NP}=-\Cc{10^\prime\mu}$) and the scenario $(\Cc{9\mu}^{\rm NP}, \Cc{9e}^{\rm NP})$. The former is coherent with the results of the 1D ``All'' fit for $\Cc{9\mu}^{\rm NP}=-\Cc{9'\mu}$ given the possibility to accommodate an SM-like $R_K$. The latter, i.e. the  $(\Cc{9\mu}^{\rm NP}, \Cc{9e}^{\rm NP})$ scenario,
points towards the main outcome of our work, namely, the preference for an LFU contribution to $\Cc{9\ell}^{\rm NP}$ as the best description of data. One may notice in Table~\ref{tab:results2D_Marcel} that both muon and electron best-fit points are of similar size, in contrast to the situation in Ref.~\cite{Alguero:2021anc}.


\begin{table}[H] 
    \centering
   \begin{adjustbox}{width=0.8\textwidth,center=\textwidth}
\begin{tabular}{c||c|c|c||c|c|c} 
 & \multicolumn{3}{c||}{All} &  \multicolumn{3}{c}{LFUV}\\
\hline
 2D Hyp.  & Best fit  & Pull$_{\rm SM}$ & p-value & Best fit & Pull$_{\rm SM}$ & p-value\\
\hline\hline
$(\Cc{9\mu}^{\rm NP},\Cc{10\mu}^{\rm NP})$ & $(-0.78,-0.13)$ & 4.0 & 24.5\,\% & $(-0.20,+0.02)$ & 0.8 & 87.7\,\% \\
$(\Cc{9\mu}^{\rm NP},\Cc{7^{\prime}})$  & $(-0.66,+0.01)$ & 3.8 & 22.7\,\% & $(-0.22,-0.03)$ & 0.9 & 88.8\,\% \\
$(\Cc{9\mu}^{\rm NP},\Cc{9^\prime\mu})$  & $(-0.85,+0.38)$ & 4.2 & 27.6\,\% & $(-0.18,-0.06)$ & 0.8 & 87.8\,\% \\
$(\Cc{9\mu}^{\rm NP},\Cc{10^\prime\mu})$  & $(-0.78,-0.16)$ & 4.1 & 25.5\,\% & $(-0.23,-0.01)$ & 0.8 & 87.6\,\% \\ 
\hline
$(\Cc{9\mu}^{\rm NP}, \Cc{9e}^{\rm NP})$ & $(-1.15,-0.93)$ & 5.2 & 42.5\,\% & $(-3.00,-2.42)$ & 1.4 & 93.8\,\%  \\
\hline
Hyp. 1 & $(-0.69,+0.12)$ & 3.7 & 21.0\,\% & $(-0.02,+0.10)$ & 0.8 & 87.5\,\% \\
Hyp. 2 & $(-0.78,-0.10)$ & 3.5 & 20.1\,\% & $(-0.02,+0.02)$ & 0.1 & 80.9\,\% \\
Hyp. 3  & $(-0.20,+0.16)$ & 2.7 & 13.9\,\% & $(-0.09,+0.05)$ & 0.8 & 87.4\,\% \\
Hyp. 4  & $(-0.19,+0.01)$ & 2.5 & 13.1\,\% & $(-0.07,-0.03)$ & 0.8 & 87.9\,\% \\
Hyp. 5 & $(-0.81,+0.12)$ & 4.2 & 26.5\,\% & $(-0.21,-0.01)$ & 0.8 & 87.7\,\% \\
\end{tabular}
\end{adjustbox}
\caption{Experimental update only. Most prominent 2D patterns of NP in $b\to s\mu^+\mu^-$. The last five rows correspond to Hypothesis 1: $(\Cc{9\mu}^{\rm NP}=-\Cc{9^\prime\mu} , \Cc{10\mu}^{\rm NP}=\Cc{10^\prime\mu})$,  2: $(\Cc{9\mu}^{\rm NP}=-\Cc{9^\prime\mu} , \Cc{10\mu}^{\rm NP}=-\Cc{10^\prime\mu})$, 3: $(\Cc{9\mu}^{\rm NP}=-\Cc{10\mu}^{\rm NP} , \Cc{9^\prime\mu}=\Cc{10^\prime\mu}$), 4: $(\Cc{9\mu}^{\rm NP}=-\Cc{10\mu}^{\rm NP} , \Cc{9^\prime\mu}=-\Cc{10^\prime\mu})$ and 5: $(\Cc{9\mu}^{\rm NP} , \Cc{9^\prime\mu}=-\Cc{10^\prime\mu})$.}
\label{tab:results2D_Marcel}
\end{table}


\begin{table}[H] 
    \centering
    \begin{adjustbox}{width=1.\textwidth,center=\textwidth}
\begin{tabular}{c||c|c|c|c|c|c}
 & $\Cc7^{\rm NP}$ & $\Cc{9\mu}^{\rm NP}$ & $\Cc{10\mu}^{\rm NP}$ & $\Cc{7^\prime}$ & $\Cc{9^\prime \mu}$ & $\Cc{10^\prime \mu}$  \\
\hline\hline
Best fit & +0.00 & -0.98 & -0.19 & +0.01 & +0.11 & -0.16 \\ \hline
1 $\sigma$ & $[-0.01,+0.02]$ & $[-1.16,-0.78]$ & $[-0.30,-0.06]$ & $[-0.01,+0.02]$ & $[-0.24,+0.48]$ &$[-0.34,+0.04]$ 
\end{tabular}
\end{adjustbox}
\caption{Experimental update only.  1$\sigma$ confidence intervals for the NP contributions to Wilson coefficients in the 6D hypothesis allowing for NP in $b\to s\mu^+\mu^-$ operators dominant in the SM and their chirally-flipped counterparts, for the fit ``All''. The Pull$_{\rm SM}$ is $3.4\sigma$ and the p-value is $24.8\%$.}
\label{tab:Fit6D_Marcel}
\end{table}

Concerning the 6D fit in Table~\ref{tab:Fit6D_Marcel}, all the NP contributions to the Wilson coefficients are compatible with zero at $1\sigma$, apart from $\Cc{10\mu}^{\rm NP}$ (compatible at 2$\sigma$) and $\Cc{9\mu}^{\rm NP}$ (above 2$\sigma$). The allowed parameter space for $\Cc{9\mu}^{\rm NP}$ has reduced w.r.t Table 3 of Ref.~\cite{Alguero:2021anc}, due to the new $R_{K^{(*)}}$ measurement from LHCb. The allowed $1\sigma$ range for $\Cc{10\mu}^{\rm NP}$ has shifted from positive~\cite{Alguero:2021anc} to negative values. The current central values of the 6D fit yield $R_K\sim 0.92$ within this theoretical framework, as can be seen from the explicit expression of $R_K$ in Ref.~\cite{Alguero:2022wkd} (with a particular role played by the sign change in $\Cc{10\mu}^{\rm NP}$ compared to Ref.~\cite{Alguero:2021anc}).

Table~\ref{tab:Fit3Dbis_Marcel} shows the results for some prominent LFU  and LFUV scenarios corresponding to fits to the ``All'' dataset, motivated by theoretical models and introduced in previous works. The Pull$_{\rm SM}$ follow exactly the same trend of a 2.8$\sigma$ reduction as compared to Ref.~\cite{Alguero:2021anc}, like in the case of the 2D fits, with two main exceptions: Scenarios 7 and 8. These two scenarios include a small LFUV component with a difference in the coupling to leptons, being a vectorial  or a left-handed lepton current, respectively. However, Scenario 8 is unique compared to other scenarios as it can be connected with the charged anomalies in $R_{D^{(*)}}$. The long-standing degeneracy between Scenario 8 and a scenario with right-handed currents (Hypothesis 5) is now broken in favour of the LFU Scenario 8, which implies that $Q_5$~\cite{Capdevila:2016ivx} should be experimentally measured close to zero ($Q_5 \sim 0.02$, see \cite{Alguero:2022wkd}) within this scenario.

\begin{table}[H]
    \centering
    \begin{adjustbox}{width=0.8\textwidth,center=\textwidth}
\begin{tabular}{lc||c|c|c|c}
\multicolumn{2}{c||}{Scenario} & Best-fit point & 1$\sigma$ &  Pull$_{\rm SM}$ & p-value \\
\hline\hline
Scenario 0 &$\Cc{9\mu}^{\rm NP} = \Cc{9e}^{\rm NP} = \Cc{9}^{\rm U}$ & $-1.15$ & $[-1.32,-0.97]$ & 
5.4 & 41.9\%\, \\
\hline
\multirow{ 3}{*}{Scenario 5} &$\Cc{9\mu}^{\rm V}$ & $-0.94$ & $[-1.35,-0.53]$ & 
\multirow{ 3}{*}{3.7} & \multirow{ 3}{*}{23.3\,\%} \\
&$\Cc{10\mu}^{\rm V}$ & $-0.27$ & $[-0.66,+0.06]$ &  \\
&$\Cc{9}^{\rm U} = \Cc{10}^{\rm U}$ & $+0.15$ & $[-0.20,+0.54]$ &\\
\hline
\multirow{ 2}{*}{Scenario 6}&$\Cc{9\mu}^{\rm V}=-\Cc{10\mu}^{\rm V}$ & $-0.26$ & $[-0.33,-0.19]$ & 
\multirow{ 2}{*}{3.7} & \multirow{ 2}{*}{21.0\,\%} \\
&$\Cc{9}^{\rm U}=\Cc{10}^{\rm U}$ & $-0.38$ & $[-0.50,-0.25]$ &\\
\hline
\multirow{ 2}{*}{Scenario 7}&$\Cc{9\mu}^{\rm V}$ & $-0.22$ & $[-0.41,-0.03]$ & 
\multirow{ 2}{*}{5.2} & \multirow{ 2}{*}{42.5\,\% }  \\
&$\Cc{9}^{\rm U}$ & $-0.93$ & $[-1.18,-0.68]$ &\\
\hline
\multirow{ 2}{*}{Scenario 8}&$\Cc{9\mu}^{\rm V}=-\Cc{10\mu}^{\rm V}$ & $-0.10$ & $[-0.16,-0.03]$ & 
\multirow{ 2}{*}{$5.3$} & \multirow{ 2}{*}{44.2\,\%} \\
&$\Cc{9}^{\rm U}$ & $-1.07$ & $[-1.25,-0.88]$ & \\
\hline\hline
\multirow{ 2}{*}{Scenario 9}&$\Cc{9\mu}^{\rm V}=-\Cc{10\mu}^{\rm V}$ & $-0.19$ & $[-0.28,-0.11]$ & 
\multirow{ 2}{*}{2.5} & \multirow{ 2}{*}{13.1\,\%} \\
&$\Cc{10}^{\rm U}$ & $-0.02$ & $[-0.19,+0.16]$ & \\
\hline
\multirow{ 2}{*}{Scenario 10}&$\Cc{9\mu}^{\rm V}$ & $-0.63$ & $[-0.80,-0.47]$ &
\multirow{ 2}{*}{3.9} & \multirow{ 2}{*}{23.1\,\%} \\
&$\Cc{10}^{\rm U}$ & $+0.09$ & $[-0.05,+0.23]$ &\\
\hline
\multirow{ 2}{*}{Scenario 11}&$\Cc{9\mu}^{\rm V}$ & $-0.68$ & $[-0.84,-0.51]$ &
\multirow{ 2}{*}{3.8} & \multirow{ 2}{*}{23.0\,\%} \\
&$\Cc{10'}^{\rm U}$ & $-0.07$ & $[-0.20,+0.06]$ &\\
\hline
\multirow{ 2}{*}{Scenario 12}&$\Cc{9'\mu}^{\rm V}$ & $+0.24$ & $[+0.10,+0.39]$ & 
\multirow{ 2}{*}{1.3} & \multirow{ 2}{*}{8.7\,\%} \\
&$\Cc{10}^{\rm U}$ & $-0.09$ & $[-0.22,+0.04]$&\\
\hline
\multirow{ 4}{*}{Scenario 13}&$\Cc{9\mu}^{\rm V}$ & $-0.86$ & $[-1.05,-0.65]$ &
\multirow{ 4}{*}{3.8} & \multirow{ 4}{*}{26.0\,\%} \\
&$\Cc{9'\mu}^{\rm V}$ & $+0.47$ & $[+0.24,+0.73]$ & \\
&$\Cc{10}^{\rm U}$ & $+0.15$ & $[-0.03,+0.33]$ & \\
&$\Cc{10'}^{\rm U}$ & $+0.16$ & $[-0.02,+0.34]$ & \\
\end{tabular}
\end{adjustbox}
\caption{Experimental update only. Most prominent patterns for LFU and LFUV NP contributions from Fit ``All''.
Scenarios 5 to 8 were introduced in Ref.~\cite{Alguero:2018nvb}.  Scenarios 9 (motivated by 2HDMs~\cite{Crivellin:2019dun}) and 10 to 13  (motivated by $Z^\prime$ models with vector-like quarks~\cite{Bobeth:2016llm}) were introduced in Ref.~\cite{Alguero:2019ptt}. }\label{tab:Fit3Dbis_Marcel} 
\end{table}

 We have also added a new Scenario 0 to Table~\ref{tab:Fit3Dbis_Marcel} where an LFU contribution to ${\cal C}_9^{\rm NP}$ is assumed. The latest measurement of the $R_{K^{(*)}}$ observables clearly favours this scenario to explain the ``All'' dataset.  Scenario 0 is on a similar footing as Scenarios 7 and 8 regarding Pull$_{\rm SM}$ as well as p-values. 
 If $R_{K^{(*)}}$ is confirmed to be compatible with the SM in other isospin channels and/or different experimental environments, Scenario 0 may become a better fit to the data and the Pull$_{\rm SM}$ may increase significantly as compared to Scenarios 7 and 8.

\section{Full update including both experimental and theoretical new inputs } \label{sec:fullupdate}

\subsection{Theoretical framework} \label{sec:framework}

We updated the experimental inputs in the previous section and discussed briefly the impact of these new data updates on NP scenarios. However these results do not constitute the main outcome of our study, which we will provide in this section. Indeed, we perform now a complete update of both theoretical and experimental inputs. Obviously, the change of theoretical inputs shifts our SM predictions for some observables. The changes introduced in our updated theoretical framework in comparison with Sec.~\ref{sec:LHCb_global} are the following:
\begin{itemize} 
\item The $B\to K^*\ell^+\ell^-$ and $B_s \to \phi \mu^+\mu^-$ form factors at large recoil have been determined in Refs.~\cite{Gubernari:2018wyi,Gubernari:2020eft} (GKvD) using the same type of Light-Cone Sum Rules (LCSRs) used in Ref.~\cite{Khodjamirian:2010vf} (KMPW), based on $B$-meson Light-Cone Distribution Amplitudes (LCDAs). In Refs.~\cite{Gubernari:2018wyi,Gubernari:2020eft} the contributions from two- and three-particle distributions were considered up to twist four with updated hadronic inputs, leading to a significant reduction of the uncertainties for the $B\to K^*\ell^+\ell^-$ mode compared to Ref.~\cite{Khodjamirian:2010vf}. In the case of the $B_s \to \phi \mu^+\mu^-$ mode, uncertainties in Ref.~\cite{Gubernari:2020eft} are slightly larger than the previously used form factors coming from Ref.~\cite{Bharucha:2015bzk}, which are based however on the light-meson LCDAs.\footnote{We have chosen to use form factors based on B-meson LCDAs, rather than light-meson LCDAs since the agreement between central values is excellent, but B-meson LCDAs have larger uncertainties, leading to a more conservative estimation.}
We keep large- and low-recoil analyses independent and therefore, we take the GKvD results for $B\to K^*$ and $B_s \to \phi$ form factors purely from their LCSR analysis at large recoil, before they perform combined fits with lattice results at low recoil~\footnote{The $B\to K^*$ form factors are parametrised through a $z$-expansion~\cite{Gubernari:2018wyi}. In the $B_s\to\phi$ case, we determined ourselves the parameters of the corresponding $z$-expansions from the values of the form factors for a few $q^2$-values given in Ref.~\cite{Gubernari:2020eft}. We checked that these parameters are in good agreement with the ones available in the EOS software~\cite{EOSAuthors:2021xpv}.}. We split the form factors into universal soft form factors and power corrections following the same approach of Refs.~\cite{Descotes-Genon:2014uoa,Capdevila:2017ert}, which allows us to take into account the correlations among the form factors not only from the LCSR outcomes but also from the relationships derived in the heavy-quark and soft-collinear effective theories~\cite{Beneke:2001at,Charles:1998dr}.
\item The $B \to K\ell^+\ell^-$ form factors have been obtained by the HPQCD collaboration using the highly improved staggered quark formalism with MILC gluon field configurations~\cite{Parrott:2022rgu,Parrott:2022dnu}. We update our form factors accordingly, replacing our previous inputs~\cite{Khodjamirian:2010vf,Bouchard:2013eph}.
\end{itemize}

This new theoretical framework supersedes the previous one used in particular in Ref.~\cite{Alguero:2021anc}. The new set of theoretical predictions for the relevant observables within the SM case are presented in App.~\ref{app:SMpredictions}.   The main changes that we observe for the SM predictions are the following:

\begin{itemize}
\item The branching fractions in the low-$q^2$ bins for the $B\to K\ell^+\ell^-$ modes are predicted with a substantially smaller uncertainty, shrinking from $\mathcal{O}(30\%)$ to $\mathcal{O}(10\%)$ without significant shifts in the central values. These new SM predictions lead to a consistent increase in the tension with experimental results for $B\to K\mu^+\mu^-$ in both isospin modes, showing tensions up to the level of $4\sigma$.
\item The branching fractions in the low-$q^2$ bins for the $B\to K^*\ell^+\ell^-$ modes have their uncertainties reduced by half, with lower central values compatible with experimental results.
\item Similarly, non-optimised angular observables undergo a noticeable reduction in their uncertainty, while still being compatible with experimental results.
\item On the other hand, the uncertainties of optimised observables ($P_i^{(\prime)}$)  remain unchanged, and the central values shift slightly  towards the experimental values, showing a reduction in tension for the LHCb  $P_5'(B^0\to K^{0*}\mu^+\mu^-)$ measurements from 2.7$\sigma$ (2.9$\sigma$)  to 1.9$\sigma$ (1.9$\sigma$) in the $[4,6]$ ($[6,8]$) $q^2$ bin. This shift constitutes one of the most important changes of our updated theoretical framework. Notice that the uncertainties are rather conservative for two different reasons discussed at length in Refs.~\cite{Descotes-Genon:2014uoa,Capdevila:2017ert}. First, our framework separating universal form factors and power corrections yields larger uncertainties for observables like $P_5^\prime$ than taking directly the GKvD form factors \cite{Gubernari:2018wyi,Gubernari:2020eft}. Second, we adopt a conservative approach for the evaluation of long-distance charm-loop contributions that allows the interference between short and long-distance to be either constructive or destructive.

\item In the case of the $B_s\to \phi\ell^+\ell^-$ modes, we see an enhancement of the uncertainties for the branching fractions with practically unchanged central values.
\end{itemize}

\subsection{Global fit results}
\label{sec:LHCb_global_results}

With this updated theoretical framework 
and the new data presented in the previous section,
we can repeat 
 our analysis for 1D, 2D, 6D and LFUV fits 
for $b \to s \ell^+\ell^-$ modes. 
This analysis presented here and performed using the new theoretical framework and new data supersedes our previous analysis~\cite{Alguero:2021anc}. The main numerical results are provided in \cref{tab:results1D,tab:results2D,tab:Fit6D,tab:Fit6DLFU,Fit3Dbismain}.


\begin{table*}[!ht] 
    \centering
    \begin{adjustbox}{width=1.\textwidth,center=\textwidth}
\begin{tabular}{c||c|c|c|c||c|c|c|c} 
 & \multicolumn{4}{c||}{All} &  \multicolumn{4}{c}{LFUV}\\
\hline
1D Hyp.   & Best fit& 1$\sigma$/2$\sigma$   & Pull$_{\rm SM}$ & p-value & Best fit & 1$\sigma$/ 2$\sigma$  & Pull$_{\rm SM}$ & p-value\\
\hline\hline
\multirow{2}{*}{$\Cc{9\mu}^{\rm NP}$}    & \multirow{2}{*}{-0.67} &    $[-0.82,-0.52]$ &    \multirow{2}{*}{4.5}   & \multirow{2}{*}{20.2\,\%}
&   \multirow{2}{*}{-0.21}   &$[-0.38,-0.04]$&   \multirow{2}{*}{1.2}  & \multirow{2}{*}{92.4\,\%}  \\
 &  & $[-0.98,-0.37]$ &  & &  &  $[-0.57,+0.12]$ & \\
 
 \multirow{2}{*}{$\Cc{9\mu}^{\rm NP}=-\Cc{10\mu}^{\rm NP}$}    &   \multirow{2}{*}{-0.19} &    $[-0.25,-0.13]$ &   \multirow{2}{*}{3.1}  & \multirow{2}{*}{9.9\,\%}
 &  \multirow{2}{*}{-0.08}   &   $[-0.15,-0.01]$ & \multirow{2}{*}{1.1}   & \multirow{2}{*}{91.6\,\%}  \\
 &  & $[-0.32,-0.07]$ &  & & & $[-0.22,+0.06]$  &    \\
 
 \multirow{2}{*}{$\Cc{9\mu}^{\rm NP}=-\Cc{9'\mu}$}     & \multirow{2}{*}{-0.47} &    $[-0.66,-0.30]$   &  \multirow{2}{*}{3.0}  & \multirow{2}{*}{9.5\,\%}
 &  \multirow{2}{*}{-0.04}   &    $[-0.26,+0.15]$  & \multirow{2}{*}{0.2} & \multirow{2}{*}{87.5\,\%} \\
 &  & $[-0.85,-0.14]$ &  & & & $[-0.52,+0.32]$ &    \\
 
\end{tabular} \end{adjustbox}
\caption{Full update. Most prominent 1D patterns of NP in $b\to s\mu^+\mu^-$. Pull$_{\rm SM}$ is quoted in units of standard deviation.  The p-value of the SM hypothesis is $5.1\%$ for the fit ``All'' and $90.2\%$ for the fit LFUV.} 
\label{tab:results1D}
\end{table*}

\begin{table*}[!ht] 
    \centering
   \begin{adjustbox}{width=0.8\textwidth,center=\textwidth}
\begin{tabular}{c||c|c|c||c|c|c} 
 & \multicolumn{3}{c||}{All} &  \multicolumn{3}{c}{LFUV}\\
\hline
 2D Hyp.  & Best fit  & Pull$_{\rm SM}$ & p-value & Best fit & Pull$_{\rm SM}$ & p-value\\
\hline\hline
$(\Cc{9\mu}^{\rm NP},\Cc{10\mu}^{\rm NP})$			&$(-0.82, -0.17)$&4.4&$21.9\%$&$(-0.26, -0.04)$&0.7&$90.1\%$\\%
$(\Cc{9\mu}^{\rm NP},\Cc{7^{\prime}})$				&$(-0.68, +0.01)$&4.2&$19.4\%$&$(-0.21, -0.02)$&0.8&$90.8\%$\\%
$(\Cc{9\mu}^{\rm NP},\Cc{9^\prime\mu})$				&$(-0.78, +0.21)$&4.3&$20.7\%$&$(-0.16, -0.08)$&0.8&$90.4\%$\\%
$(\Cc{9\mu}^{\rm NP},\Cc{10^\prime\mu})$			&$(-0.76, -0.12)$&4.3&$20.5\%$&$(-0.18, +0.04)$&0.8&$90.2\%$\\%
\hline 
$(\Cc{9\mu}^{\rm NP}, \Cc{9e}^{\rm NP})$			&$(-1.17, -0.97)$&5.6&$40.3\%$&$(-3.00, -2.45)$&1.2&$94.3\%$\\%
\hline 
Hyp. 1												&$(-0.52, +0.15)$&3.3&$12.8\%$&$(-0.01, +0.10)$&0.8&$90.2\%$\\%
Hyp. 2												&$(-0.63, -0.11)$&2.9&$10.5\%$&$(-0.04, -0.00)$&0.0&$84.1\%$\\%
Hyp. 3												&$(-0.20, +0.15)$&2.8&$9.7\%$&$(-0.09, +0.10)$&0.7&$89.7\%$\\%
Hyp. 4												&$(-0.18, -0.02)$&2.7&$9.3\%$&$(-0.06, -0.05)$&0.8&$90.2\%$\\%
Hyp. 5												&$(-0.78, +0.08)$&4.3&$20.7\%$&$(-0.16, -0.03)$&0.8&$90.3\%$\\%
\end{tabular}
\end{adjustbox}
\caption{Full update. Most prominent 2D patterns of NP in $b\to s\mu^+\mu^-$. The last five rows correspond to Hypothesis 1: $(\Cc{9\mu}^{\rm NP}=-\Cc{9^\prime\mu} , \Cc{10\mu}^{\rm NP}=\Cc{10^\prime\mu})$,  2: $(\Cc{9\mu}^{\rm NP}=-\Cc{9^\prime\mu} , \Cc{10\mu}^{\rm NP}=-\Cc{10^\prime\mu})$, 3: $(\Cc{9\mu}^{\rm NP}=-\Cc{10\mu}^{\rm NP} , \Cc{9^\prime\mu}=\Cc{10^\prime\mu}$), 4: $(\Cc{9\mu}^{\rm NP}=-\Cc{10\mu}^{\rm NP} , \Cc{9^\prime\mu}=-\Cc{10^\prime\mu})$ and 5: $(\Cc{9\mu}^{\rm NP} , \Cc{9^\prime\mu}=-\Cc{10^\prime\mu})$.}
\label{tab:results2D}
\end{table*}

\begin{table*}[!ht] 
    \centering
    \begin{adjustbox}{width=1.\textwidth,center=\textwidth}
\begin{tabular}{c||c|c|c|c|c|c}
 & $\Cc7^{\rm NP}$ & $\Cc{9\mu}^{\rm NP}$ & $\Cc{10\mu}^{\rm NP}$ & $\Cc{7^\prime}$ & $\Cc{9^\prime \mu}$ & $\Cc{10^\prime \mu}$  \\
\hline\hline
Best fit & +0.00 & -0.98 & -0.24 & +0.01 & -0.21 & -0.25 \\ \hline
1$\sigma$ & $[-0.01,+0.01]$ & $[-1.16,-0.79]$ & $[-0.35,-0.12]$ & $[-0.00,+0.03]$ & $[-0.55,+0.13]$ &$[-0.42,-0.06]$ 
\end{tabular}
\end{adjustbox}
\caption{Full update. 1$\sigma$ confidence intervals for the NP contributions to Wilson coefficients in
the 6D hypothesis allowing for NP in $b\to s\mu^+\mu^-$ operators dominant in the SM and their chirally-flipped counterparts, for the fit ``All''. The Pull$_{\rm SM}$ is $3.7\sigma$ and the p-value is $20.9\%$. }
\label{tab:Fit6D}
\end{table*}

\begin{table*}[!ht] 
    \centering
    \begin{adjustbox}{width=1.\textwidth,center=\textwidth}
\begin{tabular}{c||c|c|c|c|c|c}
 & $\Cc7^{\rm NP}$ & $\Cc{9}^{\rm U}$ & $\Cc{10}^{\rm U}$ & $\Cc{7^\prime}$ & $\Cc{9^\prime}^{\rm U}$ & $\Cc{10^\prime}^{\rm U}$  \\
\hline\hline
Best fit & +0.00 & -1.21 & +0.07 & +0.01 & -0.04 & -0.06 \\ \hline
1$\sigma$ & $[-0.01,+0.02]$ & $[-1.38,-1.03]$ & $[-0.09,+0.22]$ & $[-0.00,+0.03]$ & $[-0.40,+0.33]$ &$[-0.25,+0.13]$ 

\end{tabular}
\end{adjustbox}
\caption{Full update. 1$\sigma$ confidence intervals for the NP contributions to Wilson coefficients in
the 6D LFU hypothesis allowing for NP in $b\to s\ell^+\ell^-$ operators dominant in the SM and their chirally-flipped counterparts, for the fit ``All''. The Pull$_{\rm SM}$ is $4.6\sigma$ and the p-value is $34.0\%$. }
\label{tab:Fit6DLFU}
\end{table*}

\begin{table*}[!ht]
    \centering
    \begin{adjustbox}{width=0.8\textwidth,center=\textwidth}
\begin{tabular}{lc||c|c|c|c}
\multicolumn{2}{c||}{Scenario} & Best-fit point & 1$\sigma$ & Pull$_{\rm SM}$ & p-value \\
\hline\hline
Scenario 0 & $\Cc{9\mu}^{\rm NP} = \Cc{9e}^{\rm NP} = \Cc{9}^{\rm U}$ & $-1.17$ & $[-1.33,-1.00]$ & 
5.8 & 39.9 \%\, \\
\hline

\multirow{ 3}{*}{Scenario 5} &$\Cc{9\mu}^{\rm V}$ & $-1.02$ & $[-1.43, -0.61]$ &
\multirow{ 3}{*}{4.1} & \multirow{ 3}{*}{21.0\,\%} \\
&$\Cc{10\mu}^{\rm V}$ & $-0.35$ & $[-0.75, -0.00]$ & \\
&$\Cc{9}^{\rm U}=\Cc{10}^{\rm U}$ & $+0.19$ & $[-0.16, +0.58]$ &\\
\hline
\multirow{ 2}{*}{Scenario 6}&$\Cc{9\mu}^{\rm V}=-\Cc{10\mu}^{\rm V}$ & $-0.27$ & $[-0.34, -0.20]$ &
\multirow{ 2}{*}{4.0} & \multirow{ 2}{*}{18.0\,\%} \\
&$\Cc{9}^{\rm U}=\Cc{10}^{\rm U}$ & $-0.41$ & $[-0.53, -0.29]$ &\\
\hline
\multirow{ 2}{*}{Scenario 7}&$\Cc{9\mu}^{\rm V}$ & $-0.21$ & $[-0.39, -0.02]$ &
\multirow{ 2}{*}{5.6} & \multirow{ 2}{*}{40.3\,\% }  \\
&$\Cc{9}^{\rm U}$ & $-0.97$ & $[-1.21, -0.72]$  &\\
\hline
\multirow{ 2}{*}{Scenario 8}&$\Cc{9\mu}^{\rm V}=-\Cc{10\mu}^{\rm V}$ & $-0.08$ & $[-0.14, -0.02]$ &
\multirow{ 2}{*}{5.6} & \multirow{ 2}{*}{41.1\,\%} \\
&$\Cc{9}^{\rm U}$ & $-1.10$ & $[-1.27, -0.91]$ &\\
\hline\hline
\multirow{ 2}{*}{Scenario 9}&$\Cc{9\mu}^{\rm V}=-\Cc{10\mu}^{\rm V}$ & $-0.21$ & $[-0.29, -0.13]$ &
\multirow{ 2}{*}{2.7} & \multirow{ 2}{*}{9.3\,\%} \\
&$\Cc{10}^{\rm U}$ & $-0.06$ & $[-0.23, +0.11]$ &\\
\hline
\multirow{ 2}{*}{Scenario 10}&$\Cc{9\mu}^{\rm V}$ & $-0.65$ & $[-0.81, -0.50]$ &
\multirow{ 2}{*}{4.1} & \multirow{ 2}{*}{19.1\,\%} \\
&$\Cc{10}^{\rm U}$ & $ +0.05$ & $[-0.08, +0.18]$ &\\
\hline
\multirow{ 2}{*}{Scenario 11}&$\Cc{9\mu}^{\rm V}$ & $-0.68$ & $[-0.84, -0.52]$ &
\multirow{ 2}{*}{4.1} & \multirow{ 2}{*}{19.0\,\%} \\
&$\Cc{10'}^{\rm U}$ & $-0.03$ & $[-0.15, +0.09]$ &\\
\hline
\multirow{ 2}{*}{Scenario 12}&$\Cc{9'\mu}^{\rm V}$ & $ +0.21$ & $[+0.07, +0.34]$ &
\multirow{ 2}{*}{1.5} & \multirow{ 2}{*}{6.0\,\%} \\
&$\Cc{10}^{\rm U}$ & $-0.14$ & $[-0.26, -0.03]$ &\\
\hline
\multirow{ 4}{*}{Scenario 13}&$\Cc{9\mu}^{\rm V}$ & $-0.78$ & $[-0.97, -0.60]$ &
\multirow{ 4}{*}{3.8} & \multirow{ 4}{*}{19.2\,\%} \\
&$\Cc{9'\mu}^{\rm V}$ & $  +0.33$ & $[+0.10, +0.57]$ &\\
&$\Cc{10}^{\rm U}$ & $+0.11$ & $[-0.04, +0.26]$ &\\
&$\Cc{10'}^{\rm U}$ & $+0.13$ & $[-0.03, +0.30]$ &\\
\hline\hline

\multirow{ 2}{*}{Scenario 14}&$\Cc{9}^{\rm U}$ & $-1.16$ & $[-1.33, -0.99]$ &
\multirow{ 2}{*}{5.5} & \multirow{ 2}{*}{39.0\,\%} \\
&$\Cc{9'\mu}^{\rm V}$ & $-0.10$ & $[-0.24, +0.04]$ &\\
\hline
\multirow{ 2}{*}{Scenario 15}&$\Cc{9}^{\rm U}$ & $ -1.16$ & $[-1.33, -0.99]$ &
\multirow{ 2}{*}{5.5} & \multirow{ 2}{*}{38.4\,\%} \\
&$\Cc{10'\mu}^{\rm V}$ & $+0.03$ & $[-0.05, +0.11]$ &\\
\hline
\end{tabular}
\end{adjustbox}
\caption{Full update. Most prominent patterns for LFU and LFUV NP contributions from Fit ``All''.
Scenarios 5 to 8 were introduced in Ref.~\cite{Alguero:2018nvb}.  Scenarios 9 (motivated by 2HDMs~\cite{Crivellin:2019dun}) and 10 to 13  (motivated by $Z^\prime$ models with vector-like quarks~\cite{Bobeth:2016llm}) were introduced in Ref.~\cite{Alguero:2019ptt}. The two additional Scenarios 14 and 15 illustrate that the addition of right-handed current LFUV contributions does not improve the fit and their contribution is negligible.
}\label{Fit3Dbismain} 
\end{table*}

The new theoretical framework does not yield much difference with the results presented in Sec.~\ref{sec:LHCb_global} both for the ``All'' and the ``LFUV'' fits. 
In the case of the ``All'' fits, the SM p-value decreases from 8.0\% to 5.1\%, mainly driven by the increase of tension in $B\to K\ell^+\ell^-$ observables. The p-values of most of the NP scenarios undergo a smaller decrease, enhancing the Pull$_{\rm SM}$ of these scenarios, with the exception of the scenarios including a $\Cc{9\mu}^{\rm NP}=-\Cc{9'\mu}$ contribution (like Hyp. 1 and 2) and the $(\Cc{9\mu}^{\rm NP},\Cc{9'\mu})$ scenario, which undergo a larger decrease in p-value than the SM. 
In the case of the ``LFUV'' fits, the p-values are slightly better than in Sec.~\ref{sec:LHCb_global}, while the Pull$_{\rm SM}$ are virtually unchanged.
As could be expected given the new LHCb measurements of $R_{K(^*)}$, Scenario 0, which amounts to a lepton-flavour  universal contribution to $\Cc{9\mu}$ and $\Cc{9e}$, is particularly efficient in explaining the deviations.
We also performed 10D fits similar to Tables \ref{tab:Fit6D} and \ref{tab:Fit6DLFU} but allowing for different values for electron and muon Wilson coefficients $\Cc{9,10,9',10'\ell}$ but the results are very similar to the 6D fits for the muonic Wilson coefficients whereas the electronic Wilson coefficients are only very weakly constrained, with values compatible with LFU.

Our fully updated global fits  take the current values of $R_{K_S}$ and $R_{K^{*+}}$~\cite{LHCb:2021lvy} as we consider all the LHCb data currently available. However, the systematics 
 affecting $b\to se^+e^-$ measurements
 recently discussed by LHCb (related to missing backgrounds and particle identification) may affect not only $R_K$ and $R_{K^*}$ but also the isopin related channels $R_{K_S}$ and $R_{K^{*+}}$. In order to identify the impact of the latter in our updated results, we performed the same global fit as in Sec.~\ref{sec:fullupdate}, but without including either $R_{K_S}$ or $R_{K^{*+}}$ as data. Due to the large uncertainties on the measurements of $R_{K_S}$ or $R_{K^{*+}}$  the impact of excluding these observables is negligible for both ``All'' and ``LFUV'' fits, as we checked explicitly.

\begin{figure}
    \centering
    \includegraphics[width=7.0 cm]{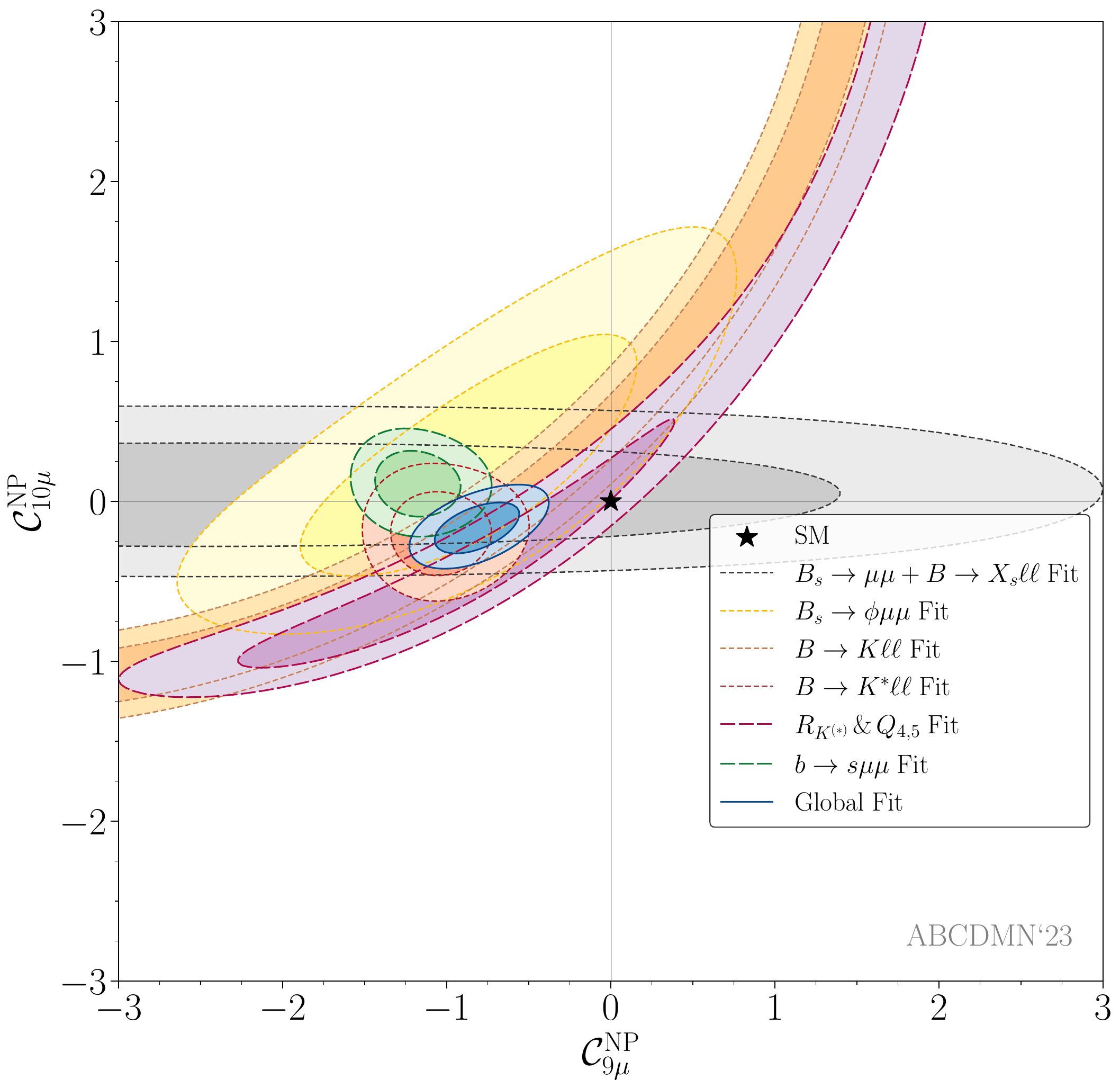} \quad
    \includegraphics[width=7.0 cm]{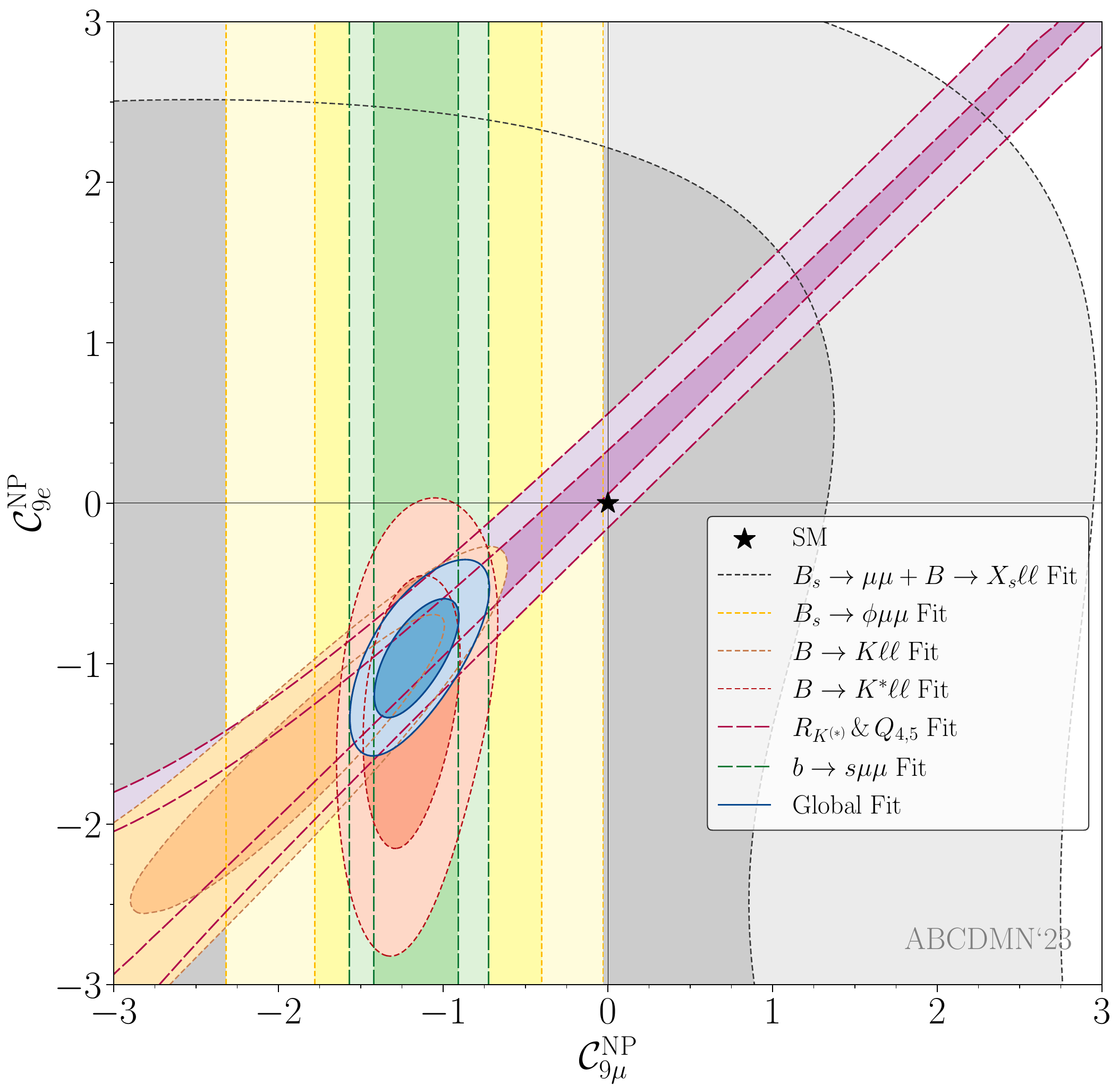}
    \caption{Full update. 1$\sigma$ (dark-shaded) and 2$\sigma$ (light-shaded) confidence regions for $(\Cc{9\mu}^{\rm NP},\Cc{10\mu}^{\rm NP})$ (left) and $(\Cc{9\mu}^{\rm NP},\Cc{9e}^{\rm NP})$ scenarios (right). Distinct fits are performed separating each of the $b\to s\ell^+\ell^-$ modes  (short-dashed contours), the LFUV observables and the  combined $b\to s\mu^+\mu^-$ modes (long-dashed contours), and the global fit (solid contours). The colour code is provided in the individual captions.  Notice that some fits (for instance the $B\to K^{(*)}\ell^+\ell^-$ Fit(s) and the LFUV Fit) share a number of observables and thus are not completely uncorrelated.}\label{fig:C9mu_C10mu}
\end{figure}
Fig.~\ref{fig:C9mu_C10mu} displays the 1 and 2$\sigma$ contours for the 2D scenarios $(\Cc{9\mu}^{\rm NP},\Cc{10\mu}^{\rm NP})$ and $(\Cc{9\mu}^{\rm NP},\Cc{9e}^{\rm NP})$ with regions corresponding to the constraints from individual modes, the LFUV observables, the combined $b\to s\mu^+\mu^-$ modes and the global fit. 

For the $(\Cc{9\mu}^{\rm NP}, \Cc{10\mu}^{\rm NP})$ scenario, the grey contour (obtained from $B(B_s\to\mu^+\mu^-)$ and $B(B\to X_s\ell^+\ell^-$)) is consistent with $\Cc{10\mu}^{\rm NP} = 0$, driven mainly by the consistency of the current global average of $B(B_s\to\mu^+\mu^-)$ with the corresponding SM estimate. While the combined $b\to s\mu^+\mu^-$ observables do prefer a slightly positive value for $\Cc{10\mu}^{\rm NP}$, the LFUV observables and the specific $B\to K^{(*)}\ell^+\ell^-$ observables prefer a more negative value, with the final outcome being that $\Cc{10\mu}^{\rm NP}$ is consistent with zero at 1$\sigma$ but has a slightly negative central value in the global fits. All the constraints are consistent at 1$\sigma$ with a value of $\Cc{9\mu}^{\rm NP} = -1$.

For the $(\Cc{9\mu}^{\rm NP},\Cc{9e}^{\rm NP})$ scenario, the effect of the new $R_{K^{(*)}}$ measurements from LHCb is visible, leading to a constraint corresponding to $\Cc{9\mu}^{\rm NP}=\Cc{9e}^{\rm NP}$ at 1$\sigma$ throughout the parameter space, hinting towards a lepton-universal NP contribution to the semileptonic $O_9$ operator. Obviously, the combination of the $b\to s\mu^+\mu^-$ modes cannot put any constraints on $\Cc{9e}^{\rm NP}$. The $B\to K\ell^+\ell^-$ observables prefer negative values for both $\Cc{9\mu,e}^{\rm NP}$ and are consistent with the relation $(\Cc{9\mu}^{\rm NP}=\Cc{9e}^{\rm NP})$ at 1$\sigma$. This stems from the fact that $R_K$ is the only $B\to K\ell^+\ell^-$ observable that contributes to $\Cc{9e}^{\rm NP}$. The $B\to K^*\ell^+\ell^-$ observables also prefer negative values for both Wilson coefficients, but with negligible correlation.

\begin{figure} 
\begin{center}
\includegraphics[width=7.0 cm]{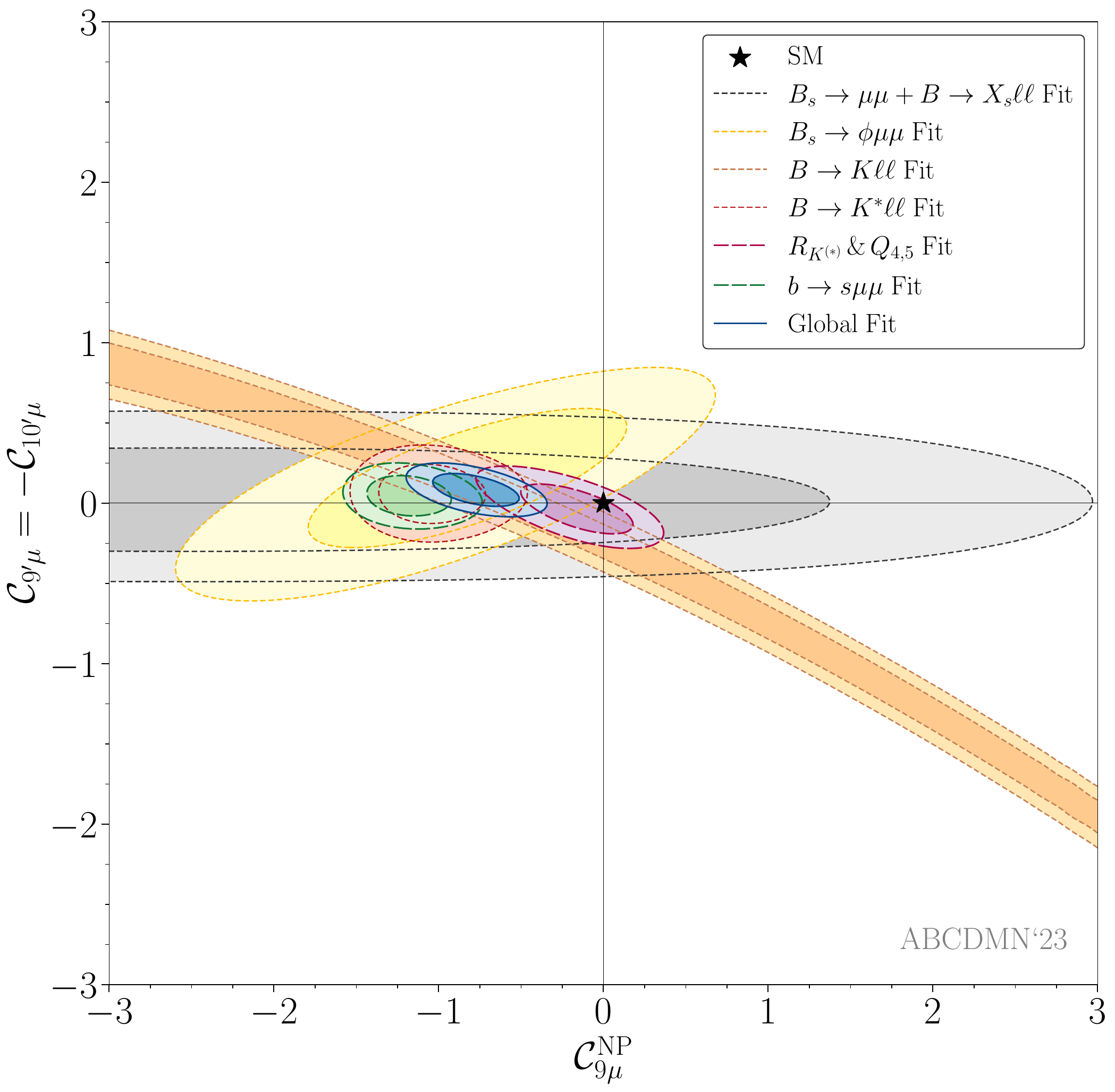} \quad
\includegraphics[width=7.0 cm]{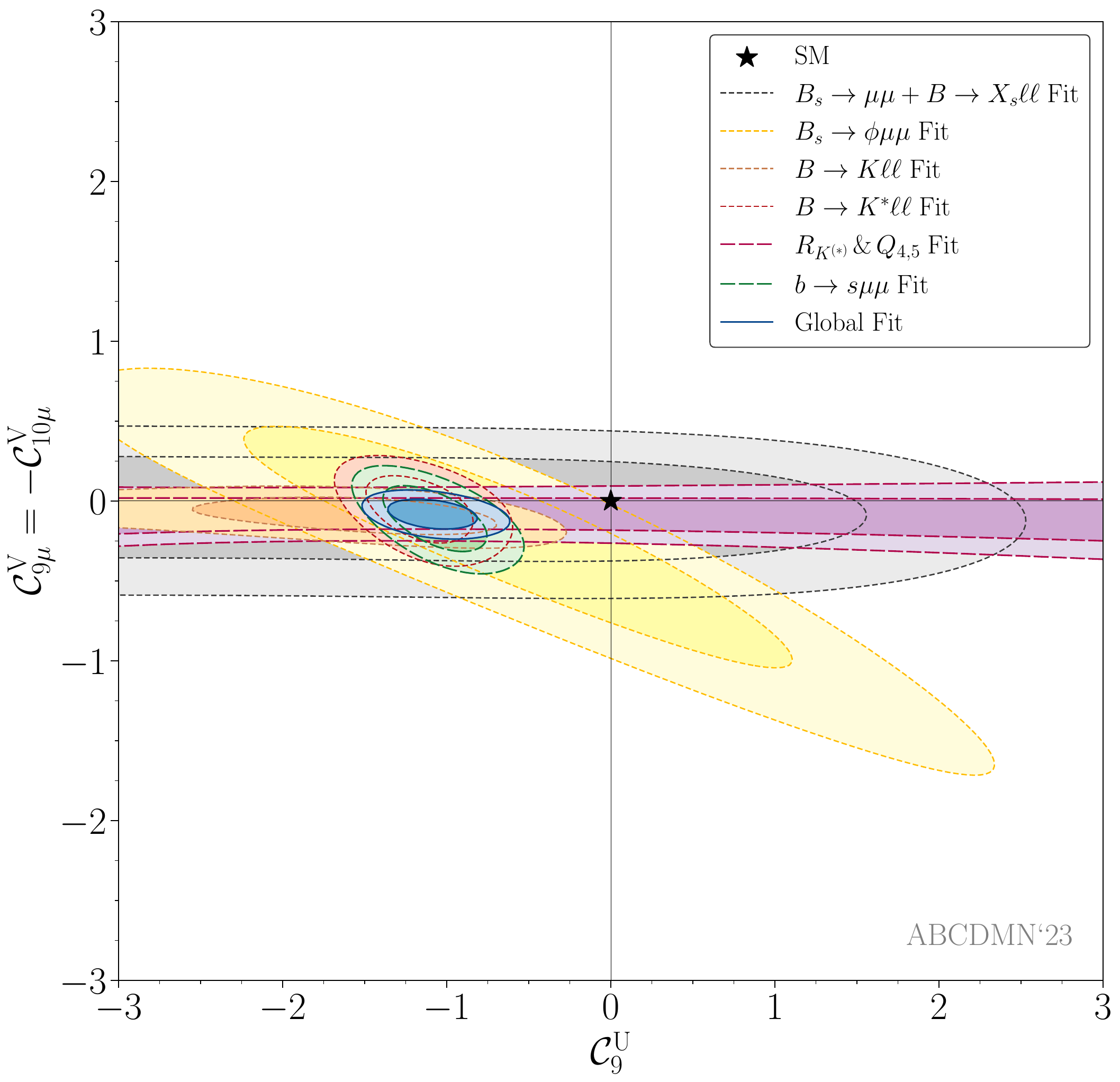}
   \end{center}  
   \caption{Full update. 1$\sigma$ (dark-shaded) and 2$\sigma$ (light-shaded) confidence intervals for the scenarios $(\Cc{9\mu}^{\rm NP}, \Cc{9^\prime\mu}=-\Cc{10^\prime\mu})$ (Hypothesis 5) to the left and $(\Cc{9}^{\rm U}, \Cc{9\mu}^{\rm V}=-\Cc{10\mu}^{\rm V})$ (Scenario 8) to the right, corresponding to the separate modes involved in the global analysis (short-dashed contours), the LFUV observables and the combined $b\to s\mu^+\mu^-$ modes (long-dashed contours) and the global fit (solid contours). The colour code is provided in the individual captions.  Notice that some fits (for instance the $B\to K^{(*)}\ell^+\ell^-$ Fit(s) and the LFUV Fit) share a number of observables and thus are not completely uncorrelated.}
   \label{fig:Hyp5_Sc8new}   
\end{figure}

\begin{figure} 
\begin{center}
\includegraphics[width=7.0 cm]{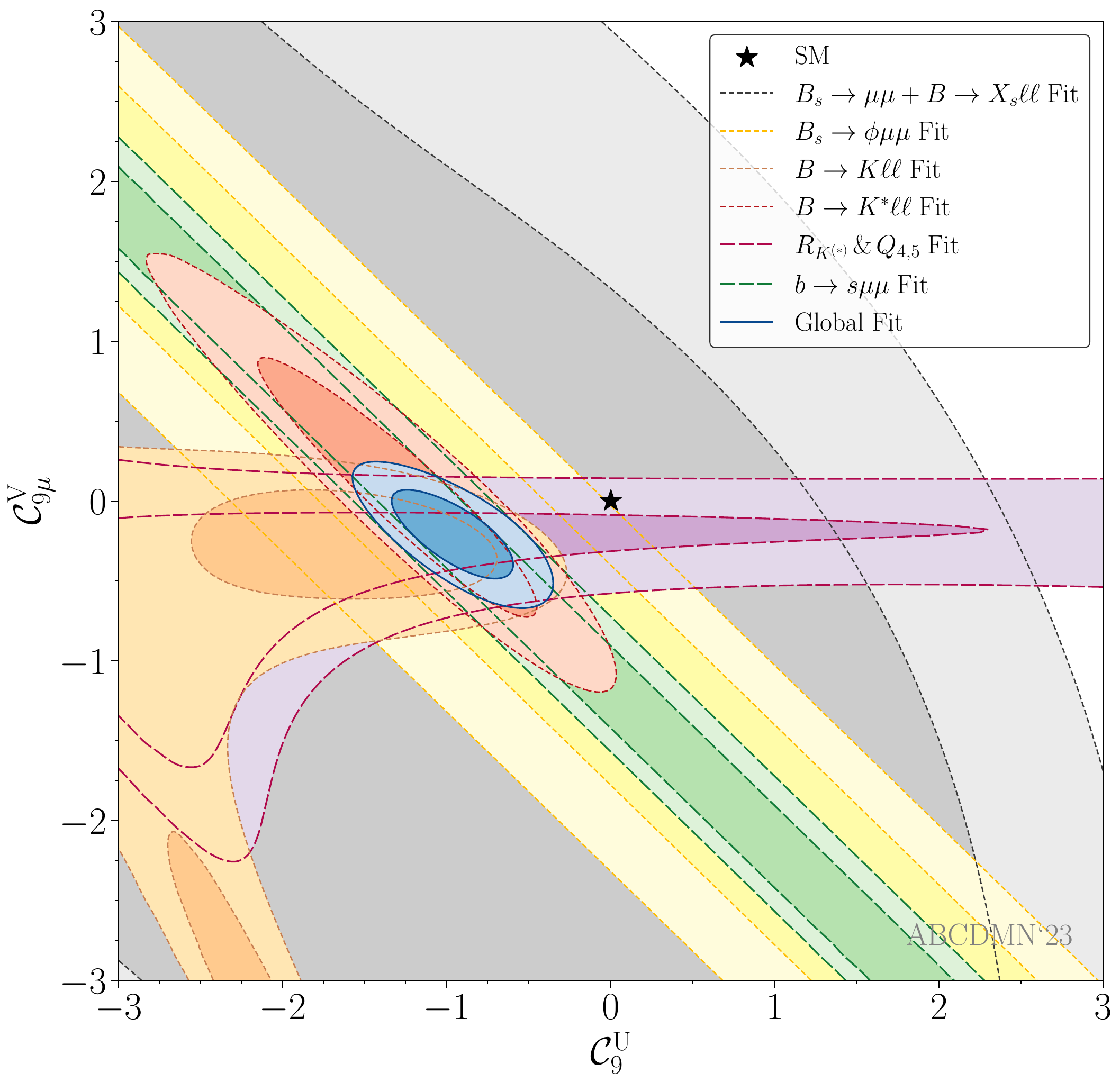} \quad
\includegraphics[width=7.0 cm]{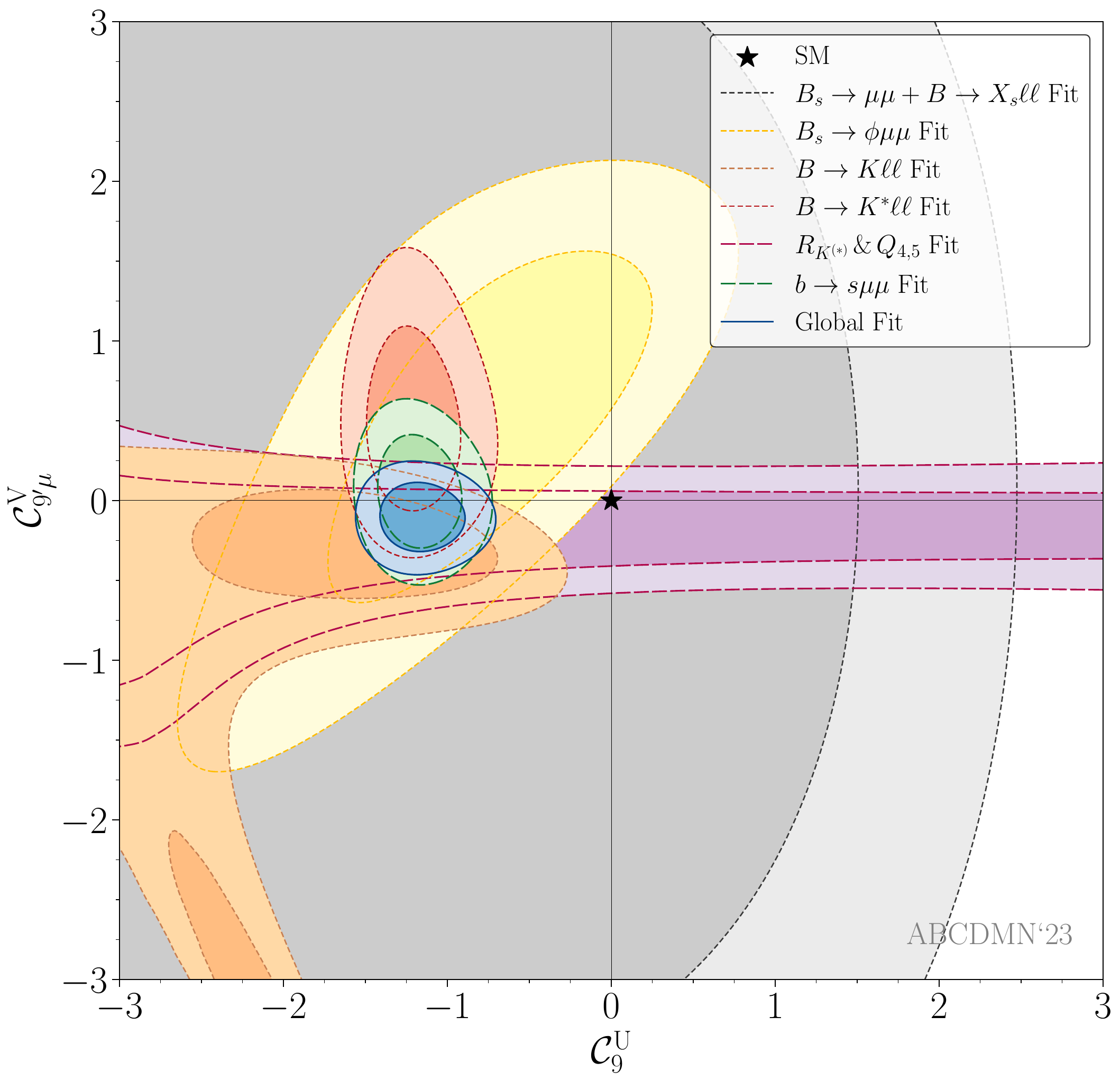}\\
\includegraphics[width=7.0 cm]{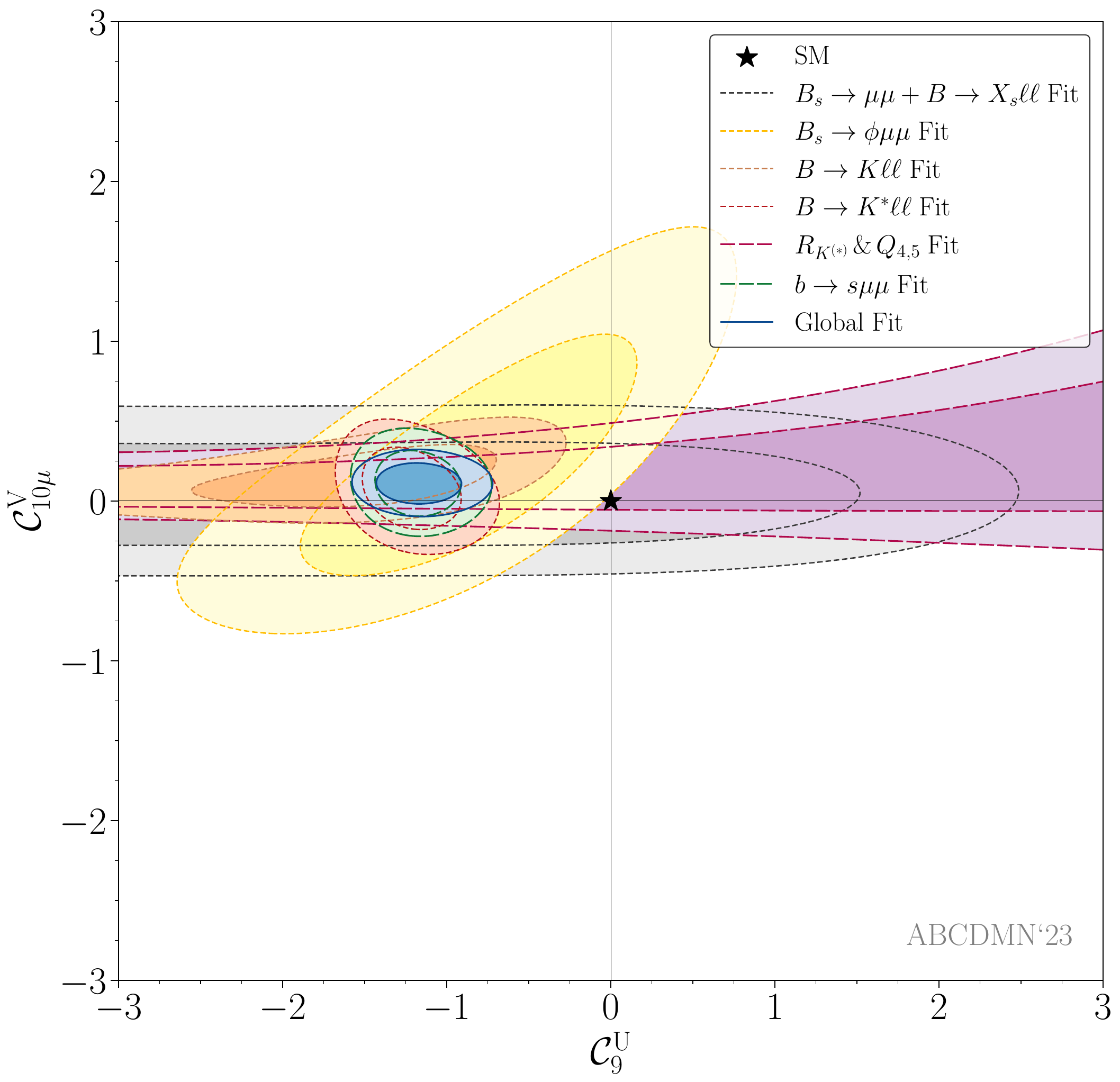}\quad
\includegraphics[width=7.0 cm]{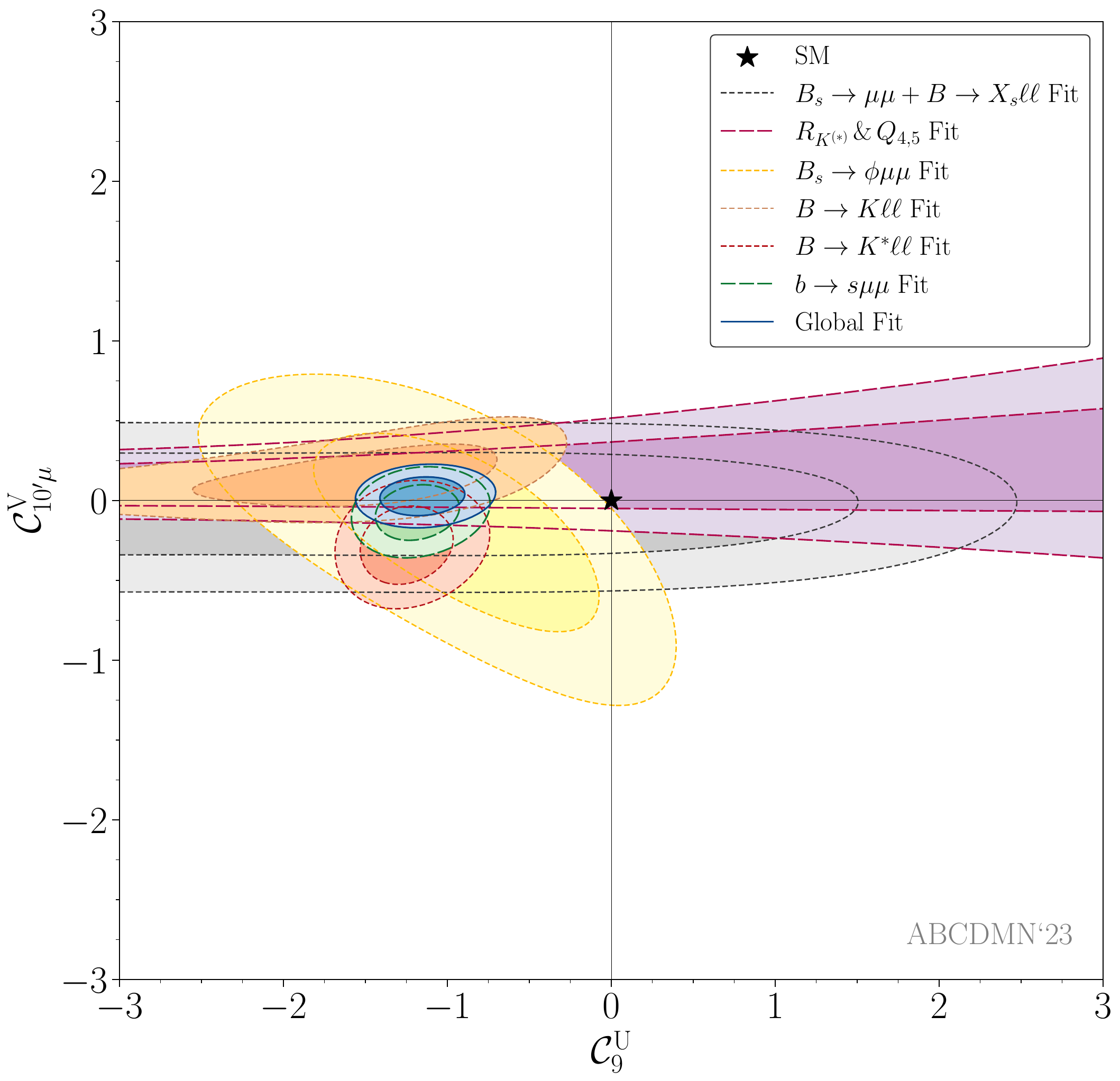}
   \end{center}  
   \caption{Full update. 1$\sigma$ (dark-shaded) and 2$\sigma$ (light-shaded) confidence intervals for the scenarios $(\Cc{9}^{\rm U}, \Cc{9\mu}^{\rm V})$ (top left), $(\Cc{9}^{\rm U}, \Cc{9^\prime\mu}^{\rm V})$ (top right), $(\Cc{9}^{\rm U}, \Cc{10\mu}^{\rm V})$ (bottom left) and $(\Cc{9}^{\rm U}, \Cc{10^\prime\mu}^{\rm V})$ (bottom right) corresponding to the separate modes involved in the global analysis (short-dashed contours), the LFUV observables and the combined $b\to s\mu^+\mu^-$ modes (long-dashed contours) and the global fit (solid contours). The colour code is provided in the individual captions. Notice that some fits (for instance the $B\to K^{(*)}\ell^+\ell^-$ Fit(s) and the LFUV Fit) share a number of observables and thus are not completely uncorrelated.}
   \label{fig:kstarkstar}   
\end{figure}

Fig.~\ref{fig:Hyp5_Sc8new} shows the 1 and 2$\sigma$ contours for the 2D cases $(\Cc{9\mu}^{\rm NP}, \Cc{9^\prime\mu}=-\Cc{10^\prime\mu})$ and $(\Cc{9}^{\rm U}, \Cc{9\mu}^{\rm V}=-\Cc{10\mu}^{\rm V})$, called Hypothesis 5 and Scenario 8 in our previous articles. In Hypothesis 5, right-handed-current contributions are compatible with vanishing values.
In Scenario 8, all constraints are consistent with $\Cc{9}^{\rm U}=-1$ at 1$\sigma$ whereas the various constraints derived in Hypothesis 5 prefer slightly different negative values for $\Cc{9\mu}^{\rm NP}$ due to the difference interference with right-handed currents for each mode, leading to a global fit consistent with $\Cc{9\mu}^{\rm NP} = -1$. The better overlap of the various constraints shows that Scenario 8 provides a better description of the data than Hypothesis 5, in agreement with the higher p-value of the former when comparing Tables~\ref{tab:results2D} and \ref{Fit3Dbismain}. 
Fig.~\ref{fig:kstarkstar} exhibits various 2D scenarios involving $\Cc{9}^{\rm U}$ with different LFUV NP contributions to Wilson coefficients. As could be expected, the new LHCb measurement of $R_{K^{(*)}}$ does not favour any such LFUV contributions, so that these scenarios end up favouring only a contribution to $\Cc{9}^{\rm U}$, for which all individual constraints overlap, in agreement with the Scenario 0 discussed at the beginning of this section.

\begin{figure} 
\begin{center}
\includegraphics[width=7.5 cm]{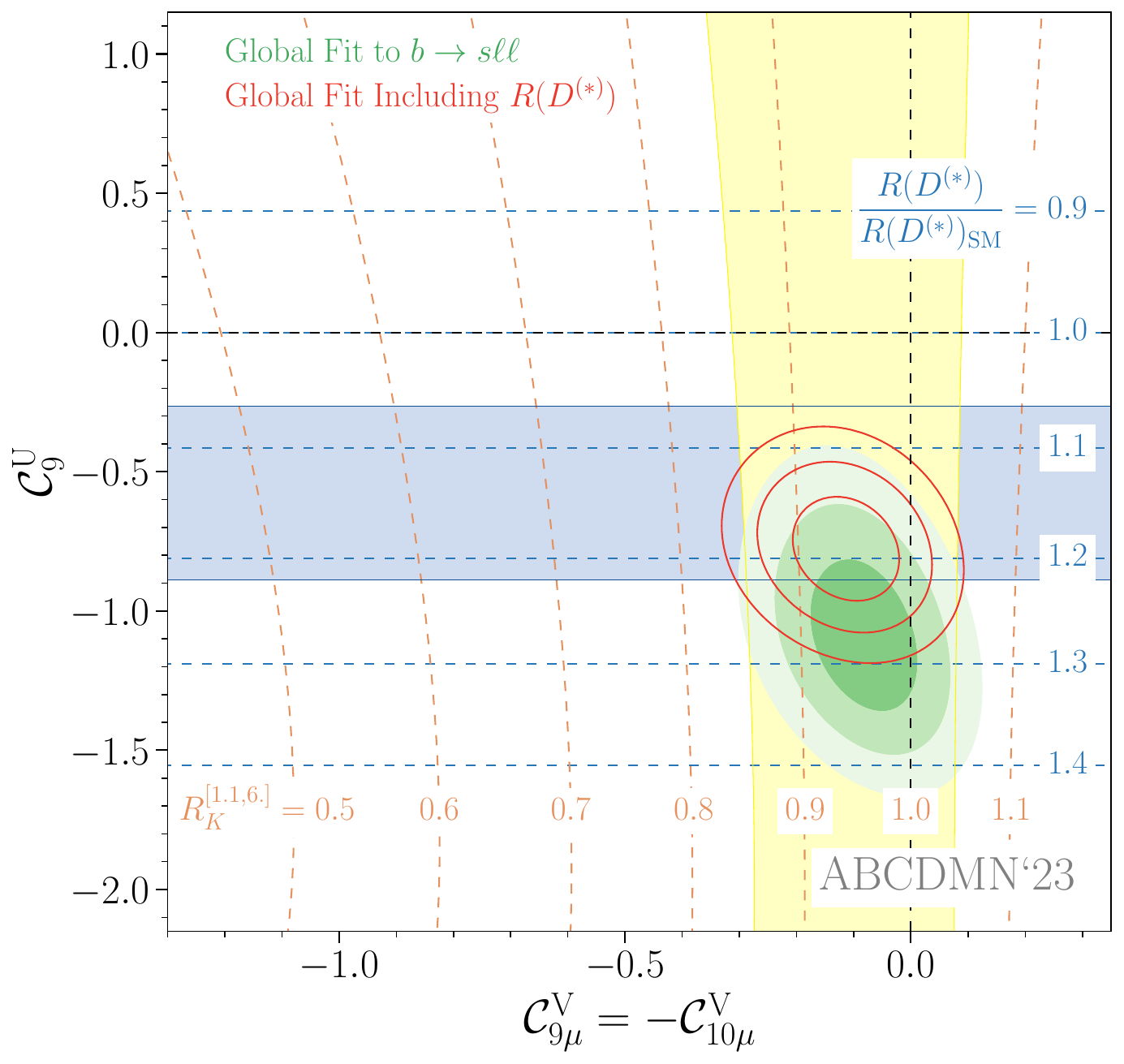}
\end{center}  
\caption{
Full update. Preferred regions at the 1, 2 and 3$\sigma$ level (green) in the  $(\Cc{9\mu}^{\rm V}=-\Cc{10\mu}^{\rm V},\Cc{9}^{\rm U})$ plane from $b\to s\ell^+\ell^-$ data. The red contour lines show the corresponding regions once $R_{D^{(\ast)}}$ is included (for a NP scale $\Lambda=2$ TeV). The horizontal blue (vertical yellow) band is consistent with $R_{D^{(\ast)}}$ ($R_K$) at the 2$\sigma$ level and the contour lines show the predicted values for these ratios.}
   \label{fig:Hyp5_Sc8RX}   
\end{figure}

Finally,  we consider Scenario 8 which provides a natural link between charged and neutral anomalies~\cite{Capdevila:2017iqn,Alguero:2019ptt}. We include the recent updates for $R_{D^{(\ast)}}$ with the fit outcome  displayed in Table~\ref{tab:scenario8imp} and Fig.~\ref{fig:Hyp5_Sc8RX}. As an indirect consequence of imposing the fulfillment of the severe bounds from $B \to K^{(*)} \nu \bar{\nu}$, a common contribution to $b \to c \tau^-\bar{\nu}_\tau$ (entering $R_{D^{(*)}}$) and $b \to s \tau^+\tau^-$ is generated~\cite{Capdevila:2017iqn}. Then through radiative effects by closing the loop of the $b \to s \tau^+\tau^-$ operator and including Renormalisation Group Evolution (RGE) effects a ${\cal C}_9^{\rm U}$ contribution is induced: 
${\cal C}_9^{\rm U}\simeq 7.5 \left(1-\sqrt{R_{D^{(*)}}/R_{D^{(*)}}^{\rm SM}} \right) \left(1+{\rm log}(\Lambda^2/(1 {\rm TeV}^2))/10.5\right)$ \cite{Alguero:2019ptt}. This link with ${\cal C}_9^{\rm U}$ is considered here (as in our previous works) for a NP scale of 2 TeV. If we take the combination of the two channels ($X=D$ and $D^* $) leading to $R_{X \, {\rm exp}}/R_{X \, {\rm SM}}=1.142 \pm 0.039$, one finds that ${\cal C}_{9}^{\rm U} \simeq -0.58$. 

Notice that the central value of ${\cal C}_{9}^{\rm U}$ 
obtained from the global fit (excluding $R_{D^{(\ast)}}$ and as shown in Table~\ref{Fit3Dbismain}), assuming this is solely realised through log-enhancement via RGE, would seem to suggest a large NP scale of few hundreds of TeV. However, such a high scale would naively imply a large suppression of the WCs, apart from also being largely incompatible with the typical NP scale that one estimates for the anomalies in the charged current $b\to c\ell\nu$.
\begin{table}[t]
 \begin{adjustbox}{width=1.\textwidth,center=\textwidth}
\begin{tabular}{|c|c|c|c|c|}%
\hline%
Scenario&Best fit&$1\sigma$&$\text{Pull}_\text{SM}$ &p-value\\%
\hline%
$(\mathcal{C}_{9\mu}^\text{V}=-\mathcal{C}_{10\mu}^\text{V},\mathcal{C}_{9}^\text{U})$&({-}0.11, {-}0.78)&({[}{-}0.17, {-}0.05{]}, {[}{-}0.90, {-}0.66{]})&6.3&$35.4\%$\\%
\hline%
\end{tabular}
\end{adjustbox}
\caption{Full update. Outcome of the fit under Scenario 8 including $R_{D^{(\ast)}}$.}
\label{tab:scenario8imp}\end{table}

\section{Investigating the role of electron modes in the global fit}
\label{sec:electronless}

\subsection{``Electron-less'' global fits}

The notorious difficulties encountered in the proper identification of electrons at LHCb have been confirmed by the significant change of the values of the LFU ratios $R_K$ and $R_{K^*}$. In this section, we attempt at apprehending the role of the LHCb measurements of electron modes and determining the impact of the rest of the data in our earlier results. We thus perform the same analysis of 1D, 2D, 6D and LFU global fits in the updated theoretical framework discussed in Sec.~\ref{sec:framework}, but with a restricted data set compared to Secs.~\ref{sec:LHCb_global} and \ref{sec:LHCb_global_results}:

\begin{itemize}
    \item We remove all LFU ratios $R_X$ (with $X=K,K_S,K^*,K^{*+}$) observables from LHCb~\cite{LHCb:2021lvy,LHCb:2022zom,LHCb:2022qnv} but keep the ones from Belle~\cite{Belle:2016fev,BELLE:2019xld,Belle:2019oag} (we thus call the fit ``electron-less'' in a slightly improper manner).
    \item We remove also all electronic observables from $b \to s e^+e^-$ modes coming from LHCb~\cite{LHCb:2020dof}. 
\end{itemize}

The full list of results in this framework can be found in \cref{app:noElectrons}. The main goal of this exercise is to gauge the consistency of the current LHCb $b \to s e^+e^-$ data with the rest of the data and see whether an improvement of the internal coherence of the fit can be reached by excluding this data.
We find several differences with respect to our previous analyses:
i) In this framework the electronic coefficients $\Cc{9e,10e}$ get only loosely constrained due to the much reduced set of electronic observables. 
ii) Since the most prominent scenarios  in the global fit correspond to scenarios with a large $\Cc{9}^{\rm U}$ contribution, the relaxation of the $\Cc{9e,10e}$ constraints does not produce an enhancement of the p-values of the best-fit scenarios, which in turn does not increase the internal consistency of the fits.
iii) The scenario $\Cc{9\mu}^{\rm NP}$ exhibits an important increase of Pull$_{\rm SM}$ up to 5.6$\sigma$ (see Table~\ref{tab:results1D_noe} in App.~\ref{app:noElectrons}) with respect to Table~\ref{tab:results1D} due to the lack of constraints from the 
LFU ratios. However, it does not provide a better fit than a universal NP contribution to $\Cc{9}$.
iv)  Scenario 9 in Table~\ref{Fit3Dbis} of App.~\ref{app:noElectrons} yields a strongly enhanced Pull$_{\rm SM}$ with respect to Table~\ref{Fit3Dbismain}, but the best-fit point corresponds to a large value of $\Cc{9\mu}^{\rm NP}$ together with 
 $\Cc{10}^{\rm U}$ and $\Cc{10\mu}^{\rm V}$ cancelling to a large extent, so that $\Cc{10\mu}^{\rm NP}$ remains small and $\Cc{10e}^{\rm NP}$ is largely unconstrained (as seen by comparing Scenario 0 in Table~\ref{Fit3Dbis} and Table~\ref{Fit3Dbismain}).

\subsection{Prediction of $R_{K}$ and $R_{K^*}$ in 
a scenario relating charged and neutral anomalies}\label{sec:predictions}

As discussed in Sec.~\ref{sec:LHCb_global_results},
 Scenario 8 provides a natural link between charged and neutral anomalies~\cite{Capdevila:2017iqn,Alguero:2019ptt}.
 If we consider the results obtained in Sec.~\ref{sec:electronless} from a data set without LHCb data on electrons, we can predict
 $R_{K^{(\ast)}}$. 
 Indeed, the indirect implications from the set of all other observables in some NP hypothesis may have consequences on observables involving electrons due to the specific structure of the NP hypothesis.

 In this scenario (${\cal C}_{9\mu}^{\rm V}=-{\cal C}_{10\mu}^{\rm V}, {\cal C}_{9}^{\rm U}$), the SM-like value for $B(B_s \to \mu^+\mu^-)$ constrains NP in the only Wilson coefficient sensitive to this branching ratio here, namely ${\cal C}_{10\mu}^{\rm V}$. The link with ${\cal C}_{9\mu}^{\rm V}$ in this scenario constrains then the size of LFUV NP. Finally, the remaining observables, mainly $P_5^{\prime\mu}$ and the link with $R_{D^{(\ast)}}$~\cite{Capdevila:2017iqn,Alguero:2019ptt}, constrain ${\cal C}_{9}^{\rm U}$. The values of the Wilson coefficients in these scenario (see Table~\ref{tab:scenario8impelectronless}) can be used to predict the values of $R_K$ and $R_{K^*}$ in each bin, leading to the values for $R_K$:
\begin{equation}
R_{K}{}_{[0.1,1.1]}=0.9018\pm0.0007\pm0.0595\,, \quad R_{K}{}_{[1.1,6.0]}=0.9086\pm0.0010\pm0.0597\,,
\end{equation}
and for $R_{K^*}$:
\begin{equation}
R_{K^*}{}_{[0.1,1.1]}=0.9606\pm0.0077\pm0.0131\,, \quad R_{K^*}{}_{[1.1,6.0]}=0.9139\pm0.0041\pm0.0523 \,,
\end{equation}

where the first error corresponds to the parametric uncertainty and the second one to the uncertainty of the Wilson Coefficient fit.
The current value of $B(B_s \to \mu^+\mu^-)$ is indeed rather constraining, meaning that the LFUV part ${\cal C}_{9\mu}^{\rm V}=-{\cal C}_{10\mu}^{\rm V}$ is small, whereas the LFU contribution ${\cal C}_{9}^{\rm U}$ is large. It is thus natural to get values of $R_K$ and $R_{K^*}$  rather compatible with LFU expectations. We can see that this scenario can accommodate LFUV in charged $b\to c\tau\nu$ decays whereas it yields LFU-only NP in $b\to s\ell^+\ell^-$, so that $R_{K^{(\ast)}}$ ratios are relatively close to 1 but optimised observables like $P_5'$ deviate from their SM predictions (in a similar way for both muonic and electronic modes).

\section{Charm-loop penguins: a mode-by-mode analysis}\label{charm}

As discussed in Sec.~\ref{sec:fullupdate}, the favoured NP scenarios correspond to situations with substantial contributions to $\Cc{9}^{\rm U}$. However, as extensively  discussed in the literature,  this effect can be difficult to distinguish from charm-loop effects. Currently, concrete computations of these effects are available using well-defined LCSR-based~\cite{Khodjamirian:2010vf,Khodjamirian:2012rm,Gubernari:2020eft} together with dispersive approaches incorporating data~\cite{Bobeth:2017vxj,Gubernari:2022hxn} showing that the impact on $B \to K\ell^+\ell^-$ is small compared to $B\to K^*\ell^+\ell^-$ and that these long-distance contributions are not large enough to accommodate the  deviations with respect to the SM measured in key observables like the optimised observable $P_5^\prime$ in $B\to K^*\mu^+\mu^-$. A recent work~\cite{Ciuchini:2022wbq} reconsidered the hypothesis of the existence of some additional unknown contributions coming from charm loops that could mimic NP contributions. This hypothesis can only become a realistic possibility and not just a speculation  if 
one is able  a) to compute these contributions 
and explain which kind of dynamical enhancement was missed in earlier computations and 
b) to prove that these contributions are universal\footnote{By universal we mean independent of the decay mode and  $q^2$-bin. Moreover, for decays to vector mesons it should also be independent of the helicity amplitude.} in order to mimic a short-distance NP contribution. We are not aware of current computations supporting this claim, whereas it is well known that a NP contribution to $\Cc{9}^{\rm U}$ is universal, while a hadronic contribution is not expected to be\footnote{While there is no proof of the inexistence of such hadronic universal contribution from first principles, there is no single example in the literature of the opposite either. Indeed all the known explicit evaluations in Refs.~\cite{Khodjamirian:2010vf,Khodjamirian:2012rm,Gubernari:2020eft} always show a non-universal behaviour of the hadronic contributions.
}. A constructive and explicit approach to make progress on this question consists in determining the preferred values for $\Cc{9}^{\rm U}$ for different $q^2$ bins and different modes, in order to check whether any variations arise, which could point towards some unknown contribution of hadronic origin responsible for these variations.

\begin{figure}
    \centering
    \includegraphics[width=0.8\textwidth]{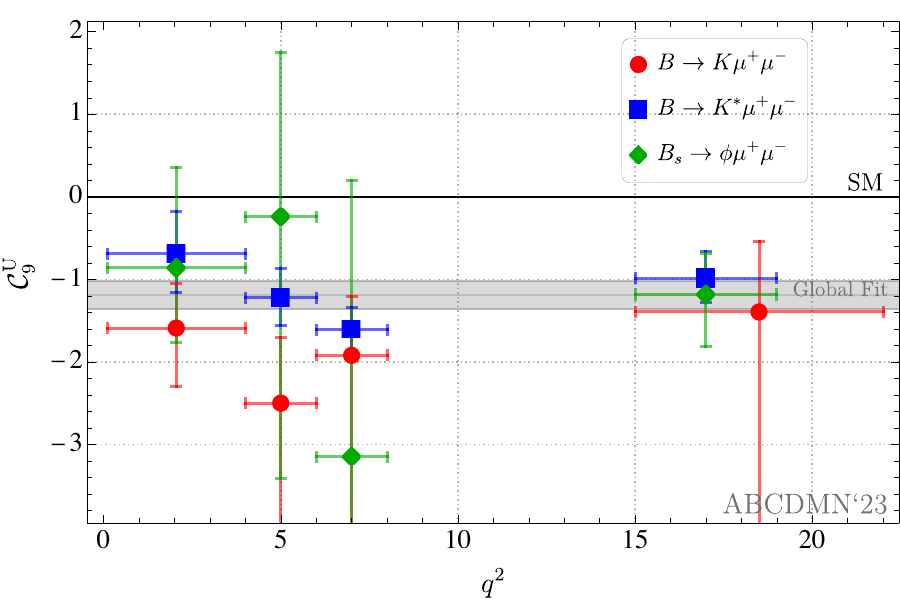}
    \caption{Preferred value of ${\cal C}^{\rm U}_{9}$ for different bins of $q^2$ for the $B\to K\mu^+\mu^-$ (red) $B\to K^*\mu^+\mu^-$ (blue) and $B_s\to \phi\mu^+\mu^-$ (green) modes. 
    The data included in these fits correspond to the branching fractions for both the charged and neutral  $B\to K^{(*)}\mu^+\mu^-$, the branching fractions for the $B_s\to \phi\mu^+\mu^-$ mode, and the angular observables for both the $B\to K^{*}\mu^+\mu^-$ and $B_s\to \phi\mu^+\mu^-$ modes. The grey band corresponds to the 1$\sigma$ confidence interval for the global fit to $\Cc{9}^{\rm U}$.}\label{fig:C9bybins}
\end{figure}

\begin{figure}
    \centering
    \includegraphics[width=0.8\textwidth]{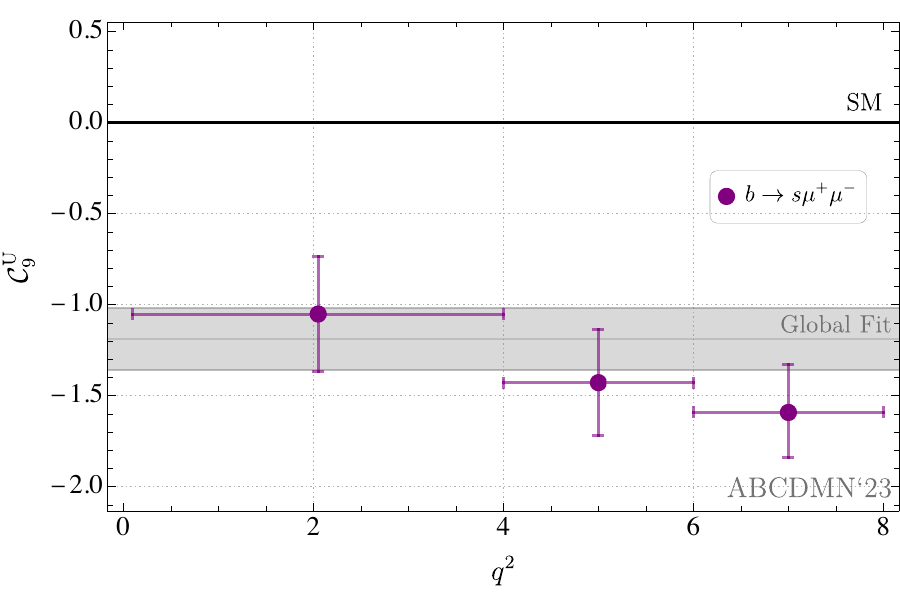}
    \caption{Preferred value of ${\cal C}^{\rm U}_{9}$ for different bins of $q^2$ in the low-$q^2$ region combining all $b\to s \mu^+\mu^-$ modes shown in \cref{fig:C9bybins}. The grey band corresponds to the 1$\sigma$ confidence interval for the global fit to $\Cc{9}^{\rm U}$. }\label{fig:C9bybinscombined}
\end{figure}

\begin{figure}
    \centering
    \includegraphics[width=0.8\textwidth]{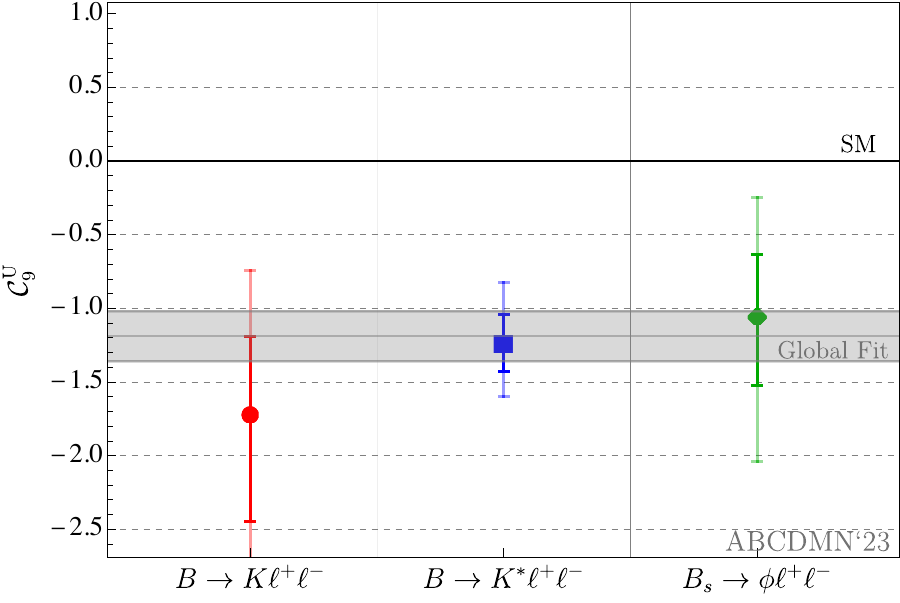}
    \caption{Preferred values of ${\cal C}^{\rm U}_{9}$, and their 1$\sigma$ and 2$\sigma$ confidence intervals, for the different $b\to s\ell^+\ell^-$ modes. The grey band corresponds to the 1$\sigma$ confidence interval for the global fit to $\Cc{9}^{\rm U}$. }\label{fig:C9bymodes}
\end{figure}

In \cref{fig:C9bybins} we show the best-fit points and the 1$\sigma$ intervals for $\Cc{9}^{\rm U}$ (assuming the SM value for the other Wilson coefficients) when we perform fits on distinct data sets for the $B\to K\mu^+\mu^-$, the $B\to K^*\mu^+\mu^-$  and the $B_s\to \phi\mu^+\mu^-$ modes in different $q^2$-bins~\footnote{We combined the available bins inside the $[0.1,4.0]$ interval as  different modes and experiments performed measurements in different 
bins in this energy, preventing a direct comparison in narrower bins.}. In all but two cases the fit results are compatible within 1$\sigma$ with the range for $\Cc{9}^{\rm U}$ from Scenario 0. The remaining two cases are compatible well within 2$\sigma$, with preferred values slightly more negative than the 
global fit results.
A potential energy and mode dependence of $\Cc{9}^{\rm U}$ can be further discussed by looking at 
\cref{fig:C9bybinscombined,fig:C9bymodes} respectively, which show an excellent agreement between the preferred values  of $\Cc{9}^{\rm U}$ obtained by fitting each of the available $b\to s\mu^+\mu^-$ modes independently, and each of the available bins independently.   
Obviously, this test requires more accurate experimental inputs to be fully conclusive, but it provides direct support from the data to the short-distance source of $\Cc{9}^{\rm U}$.

An alternative way to investigate the nature of the $\Cc{9}^{\rm U}$ contribution consists in identifying NP in other modes which can generate (through RGE) a NP contribution to  $\Cc{9}^{\rm U}$, as described in detail in Ref.~\cite{Alguero:2022wkd}. A particular prominent
candidate is given by NP in $b \to s \tau^+\tau^-$, which can be connected to anomalies in $b\to c\tau\nu$ decays and generate a universal contribution to $b\to s\ell^+\ell^-$ through RGE~\cite{Capdevila:2017iqn, Cornella:2020aoq, Alguero:2022wkd}.

\section{Conclusions} \label{sec:conclusions}

Over the last decade, a set of intringuing discrepancies with respect to SM expectations have been noticed in $b\to s\mu^+\mu^-$ and $b\to c\tau\nu$ decays by the LHCb collaboration. In this context, hints of LFU violation have been discussed when comparing $b\to se^+e^-$ and $b\to s\mu^+\mu^-$ decays. Recent results from the LHCb collaboration~\cite{LHCb:2022zom,LHCb:2022qnv} with a full data set of 9 fb$^{-1}$ indicate that the LFU ratios $R_K$ and $R_{K^*}$ are in excellent agreement with SM expectations and that
previous hints of deviations stem actually from a combination of statistical fluctuations and systematic effects affecting electron modes. Let us stress that the deviations observed both in $b\to s\mu^+\mu^-$ and $b\to c\tau\nu$ decays remain and require an interpretation, whether in terms of experimental effects, theoretical systematics or NP explanations.

We therefore reconsider our model-independent analysis of NP scenarios in order to determine whether these scenarios are still favoured in this new landscape. We perform this analysis relying on the same general framework as in our earlier works (c.f. \cite{Alguero:2021anc} and references therein). In order to identify the impact of each component of the analysis, we consider three different cases: a partial update concerning only the experimental data and in particular the new $R_K$ and $R_{K^*}$ data, a full update concerning both experimental and theoretical inputs, and an analysis without LHCb results on electron modes.

Most scenarios and hypotheses exhibit a reduction of the Pull$_{\rm SM}$ between 2 to 3$\sigma$ when we consider both partial and full updates. In particular the LFUV fits remain below the 1.5$\sigma$ level of significance for our main update. Our new theoretical framework does not yield significant differences in terms of Pull$_{\rm SM}$ and hierarchy of hypotheses compared to our previous framework. There are however differences in terms of pulls for some observables.
We can see that the tension for $B(B\to K\mu^+\mu^-)$ increases due to the new determination of the form factors based on lattice QCD. The tension for $P_5^\prime$ in the two anomalous bins decreases to $\sim 2\sigma$, the reason being that the use of updated form factors determined from LCSR slightly shifts central value predictions  towards the SM.
However, even if the new form factors have a reduced uncertainty, the total uncertainty of this observable remains essentially the same as it is dominated by factorizable power corrections, charm-loop long-distance contributions and parametric uncertainties in our approach.



We may highlight four specific conclusions of our studies.
i) The comparison of ${\cal C}_{9\mu}^{\rm NP}$ and ${\cal C}_{9\mu}^{\rm NP}=-{\cal C}_{10\mu}^{\rm NP}$ scenarios is now clearly favouring the former, due to the marginal space left for NP in ${\cal C}_{10\mu}$ by the new world average for $B(B_s \to \mu^+\mu^-)$. ii) There is now also a  clear preference for a scenario with LFU-NP (Scenario 8) versus a scenario with right-handed currents (like Hypothesis 5). iii) The scenario with LFU NP in ${\cal C}_9$ seems now an optimal option to explain $b\to s\ell^+\ell^-$ and $b\to s\gamma$ data. iv) Scenario 8 (${\cal C}_{9\mu}^{\rm V}=-{\cal C}_{10\mu}^{\rm V}, {\cal C}_{9}^{\rm U}$) remains a viable option to accommodate the charged anomalies in $R_{D^{(\ast)}}$.

We also discussed the impact of removing all LHCb electronic observables from the fit. The electronic Wilson coefficients are then only loosely constrained but there is no significant increase in the p-value of the fits and thus no enhancement of the internal consistency of the fit. We also predict the value of $R_K$ and $R_{K^*}$ within a specific scenario tying neutral ($b\to s\ell^+\ell^-$) and charged ($b\to c\tau\nu$) anomalies. We can see that the current data is compatible with this scenario where lepton-flavour-universality violating NP in $b\to c\tau\nu$ decays (affecting $R_{D^{(\ast)}}$) yields lepton-flavour-universal contributions in $b\to s\ell^+\ell^-$ through RGE (keeping $R_{K^{(\ast)}}$ at the SM value).

The mode-by-mode and bin-by-bin analysis of the $\Cc{9\mu}$ contributions shows a consistency between the different  $b\to s\mu^+\mu^-$ modes, and furthermore, no significant $q^2$ dependence is found. This
is fully compatible with a short-distance origin of the shift in $\Cc{9}^{\rm U}$, i.e. an LFU NP contribution.
A hadronic explanation should not only require the contributions to be substantially larger than the estimations present in the literature, but also be primarily mode, helicity and $q^2$ independent. This goes against natural expectations concerning hadronic contributions and would require a specific mechanism to explain both the enhanced size and the independence of the mode and the energy.

The situation after the recent LHCb results on LFU violation is thus complicated to ascertain: deviations in $b\to s\mu^+\mu^-$ and $b\to c\tau\nu$ remain and still support some NP models, but the disappearance of the cleanest deviations make their interpretation more dependent on our understanding of the hadronic dynamics of these modes. Fortunately, recent progress has been made in this direction, with new determination of form factors and a better understanding of non-local contributions. But the main contribution in the following years will come from Belle II: time will tell if they can provide some additional support to the deviations observed by the LHCb experiment and thus confirm that these decays are a credible path to NP.

\section*{Acknowledgements}

This project has received support from the European Union’s Horizon 2020 research and innovation programme under the Marie Sklodowska-Curie grant agreement No 860881-HIDDeN [S.D.G.]. This study has been carried out within the INFN project (Iniziativa
Specifica) QFT-HEP [M.N.B.]. J.M. gratefully acknowledges the financial support from ICREA under the ICREA Academia programme and from the Pauli Center (Zurich) and the Physics Department of University of Zurich. J.M. and A.B. also received financial support from Spanish Ministry of Science, Innovation and Universities (project PID2020-112965GB-I00) and from the Research Grant Agency of the Government of Catalonia (project SGR 1069). The work of B.C. is supported by the Margarita Salas postdoctoral program funded by the European Union-NextGenerationEU. B.C. also thanks the CERN Theory Division, where part of this work was done, for hospitality and the Cambridge Pheno Working Group for helpful discussions.

\appendix



\section{Global fits without LHCb data on electrons} \label{app:noElectrons}

We present the full list of results related to fits presented in Sec.~\ref{sec:electronless}. We perform the fits  in the updated theoretical framework discussed in Sec.~\ref{sec:framework}, with a restricted data set compared to Secs.~\ref{sec:LHCb_global} and \ref{sec:LHCb_global_results}, removing all LHCb electron-related results as described in Sec.~\ref{sec:electronless}. The main objective of this exercise is to gauge the coherence of the current LHCb $b \to s e^+e^-$ data with the rest of the dataset and see whether an improvement of the internal coherence of the fit can be reached by excluding this data.

\begin{table}[H] \scriptsize
    \centering
    \begin{adjustbox}{width=0.7\textwidth,center=\textwidth}
\label{sec:Global1DFits1}%
\begin{tabular}[H]{c||c|c|c|c}%
Scenario&Best fit&$1\sigma$/$2\sigma$&$\text{Pull}_\text{SM}$ &p-value\\%
\hline\hline
\multirow{2}{*}{$\mathcal{C}_{9\mu}^\text{NP}$}&\multirow{2}{*}{-1.12}&{[}{-}1.29, {-}0.94{]}&\multirow{2}{*}{5.6}&\multirow{2}{*}{29.1\%}\\
& & {[}{-}1.44, {-}0.75{]}& &\\ 
\multirow{2}{*}{$\mathcal{C}_{9\mu}^\text{NP}=-\mathcal{C}_{10\mu}^\text{NP}$}&\multirow{2}{*}{{-}0.37}&{[}{-}0.48, {-}0.26{]}&\multirow{2}{*}{3.5}&\multirow{2}{*}{9.2\%}\\
& &{[}{-}0.60, {-}0.15{]} & &\\
\multirow{2}{*}{$\mathcal{C}_{9\mu}^\text{NP}=-\mathcal{C}_{9'\mu}$}&
\multirow{2}{*}{{{-}0.73}}&{[}{-}0.91, {-}0.53{]}&\multirow{2}{*}{3.8}&\multirow{2}{*}{10.4\%}\\
& &{[}{-}1.08, {-}0.34{]} & &\\
\hline%
\end{tabular}
\end{adjustbox}
\caption{``Electron-less'' fits. Most prominent 1D patterns of NP in $b\to s\ell^+\ell^-$ excluding all electronic observables measured by LHCb.  The p-value of the SM hypothesis is $3.6\%$ for the ``Electron-less'' fit.}
\label{tab:results1D_noe}
\end{table}

\begin{table}[H] 
    \centering
    \begin{adjustbox}{width=0.6\textwidth,center=\textwidth}
\label{sec:Global1DFits2}%
\begin{tabular}[H]{c||c|c|c}%


Scenario&Best fit&$\text{Pull}_\text{SM}$ &p-value \\%
\hline\hline
$(\mathcal{C}_{9\mu}^\text{NP},\mathcal{C}_{10\mu}^\text{NP})$&({-}1.12, +0.06)&5.2&27.9\%\\%
$(\mathcal{C}_{9\mu}^\text{NP},\mathcal{C}_{7'})$&({-}1.13, +0.02)&5.3&29.1\%\\%
$(\mathcal{C}_{9\mu}^\text{NP},\mathcal{C}_{9'\mu})$&({-}1.13, +0.10)&5.2&27.9\%\\%
$(\mathcal{C}_{9\mu}^\text{NP},\mathcal{C}_{10'\mu})$&({-}1.14, {-}0.09)&5.3&28.6\%\\%
\hline%
$(\mathcal{C}_{9\mu}^\text{NP},\mathcal{C}_{9e}^\text{NP})$&({-}1.17, {-}1.29)&5.5&32.3\%\\%
\hline%
Hyp. 1&({-}0.94, +0.32)&4.3&16.6\%\\%
Hyp. 2&({-}0.76, {-}0.07)&3.5&10.2\%\\%
Hyp. 3&({-}0.43, +0.28)&3.4&10.0\%\\%
Hyp. 4&({-}0.37, {-}0.07)&3.2&8.9\%\\%
Hyp. 5&({-}1.14, +0.06)&5.2&28.4\%\\%

\end{tabular}
\end{adjustbox}
\caption{``Electron-less'' fits. Most prominent 2D patterns of NP in $b\to s\ell^+\ell^-$ excluding all electronic observables measured by LHCb. The last five rows correspond to Hypothesis 1: $(\Cc{9\mu}^{\rm NP}=-\Cc{9^\prime\mu} , \Cc{10\mu}^{\rm NP}=\Cc{10^\prime\mu})$,  2: $(\Cc{9\mu}^{\rm NP}=-\Cc{9^\prime\mu} , \Cc{10\mu}^{\rm NP}=-\Cc{10^\prime\mu})$, 3: $(\Cc{9\mu}^{\rm NP}=-\Cc{10\mu}^{\rm NP} , \Cc{9^\prime\mu}=\Cc{10^\prime\mu}$), 4: $(\Cc{9\mu}^{\rm NP}=-\Cc{10\mu}^{\rm NP} , \Cc{9^\prime\mu}=-\Cc{10^\prime\mu})$ and 5: $(\Cc{9\mu}^{\rm NP} , \Cc{9^\prime\mu}=-\Cc{10^\prime\mu})$.}
\label{tab:results2D_noe}
\end{table}
\begin{table}[H] 
    \centering
    \begin{adjustbox}{width=1.\textwidth,center=\textwidth}
\begin{tabular}{c||c|c|c|c|c|c}
 & $\Cc7^{\rm NP}$ & $\Cc{9\mu}^{\rm NP}$ & $\Cc{10\mu}^{\rm NP}$ & $\Cc{7^\prime}$ & $\Cc{9^\prime \mu}$ & $\Cc{10^\prime \mu}$  \\
\hline\hline
Best fit & +0.00 & -1.17 & +0.01 & +0.02 & -0.09 & -0.11 \\ \hline
1$\sigma$ & $[-0.01,+0.02]$ & $[-1.34,-0.98]$ & $[-0.14, +0.16]$ & $[+0.00,+0.03]$ & $[-0.45, +0.28]$ &$[-0.29, +0.08]$ 

\end{tabular}
\end{adjustbox}
\caption{``Electron-less'' fits. Most prominent 6D patterns of NP in $b\to s\ell^+\ell^-$ excluding all electronic observables measured by LHCb. The Pull$_{\rm SM}$ is $4.4\sigma$ and the p-value is $24.2\%$.}
\label{tab:results6D_noe1}
\end{table}

\begin{table}[H]
    \centering
    \begin{adjustbox}{width=0.8\textwidth,center=\textwidth}
\begin{tabular}{lc||c|c|c|c}
\multicolumn{2}{c||}{Scenario} & Best-fit point & 1$\sigma$ & Pull$_{\rm SM}$ & p-value \\
\hline\hline
Scenario 0 & $\Cc{9\mu}^{\rm NP} = \Cc{9e}^{\rm NP} = \Cc{9}^{\rm U}$ & $-1.17$ & $[-1.33,-0.99]$ & 
5.8 & 34.5 \%\, \\
\hline

\multirow{ 3}{*}{Scenario 5} &$\Cc{9\mu}^{\rm V}$ & $-0.74$ & $[-1.59, -0.11]$ &
\multirow{ 3}{*}{4.9} & \multirow{ 3}{*}{26.8\,\%} \\
&$\Cc{10\mu}^{\rm V}$ & $ +0.44$ & $[-0.41, +1.06]$ & \\
&$\Cc{9}^{\rm U}=\Cc{10}^{\rm U}$ & $-0.38$ & $[-1.00, +0.46]$ &\\
\hline

\multirow{ 2}{*}{Scenario 6}&$\Cc{9\mu}^{\rm V}=-\Cc{10\mu}^{\rm V}$ & $-0.59$ & $[-0.70, -0.48]$ &
\multirow{ 2}{*}{5.2} & \multirow{ 2}{*}{28.2\,\%} \\
&$\Cc{9}^{\rm U}=\Cc{10}^{\rm U}$ & $-0.53$ & $[-0.64, -0.42]$ &\\
\hline

\multirow{ 2}{*}{Scenario 7}&$\Cc{9\mu}^{\rm V}$ & $+0.13$ & $[-0.55, +0.72]$ &
\multirow{ 2}{*}{5.5} & \multirow{ 2}{*}{32.3\,\%}  \\
&$\Cc{9}^{\rm U}$ & $-1.29$ & $[-1.90, -0.61]$  &\\
\hline
\multirow{ 2}{*}{Scenario 8}&$\Cc{9\mu}^{\rm V}=-\Cc{10\mu}^{\rm V}$ & $-0.07$ & $[-0.20, +0.05]$ &
\multirow{ 2}{*}{5.5} & \multirow{ 2}{*}{32.8\,\%} \\
&$\Cc{9}^{\rm U}$ & $-1.11$ & $[-1.31, -0.90]$ &\\
\hline\hline
\multirow{ 2}{*}{Scenario 9}&$\Cc{9\mu}^{\rm V}=-\Cc{10\mu}^{\rm V}$ & $-1.03$ & $[-1.20, -0.84]$ &
\multirow{ 2}{*}{4.8} & \multirow{ 2}{*}{21.9\,\%} \\
&$\Cc{10}^{\rm U}$ & $-0.94$ & $[-1.15, -0.71]$ &\\
\hline
\multirow{ 2}{*}{Scenario 10}&$\Cc{9\mu}^{\rm V}$ & $-1.12$ & $[-1.29, -0.94]$ &
\multirow{ 2}{*}{5.2} & \multirow{ 2}{*}{28.2\,\%} \\
&$\Cc{10}^{\rm U}$ & $ +0.08$ & $[-0.05, +0.22]$ &\\
\hline
\multirow{ 2}{*}{Scenario 11}&$\Cc{9\mu}^{\rm V}$ & $-1.14$ & $[-1.30, -0.96]$ &
\multirow{ 2}{*}{5.2} & \multirow{ 2}{*}{28.2\,\%} \\
&$\Cc{10'}^{\rm U}$ & $-0.07$ & $[-0.19, +0.05]$ &\\
\hline
\multirow{ 2}{*}{Scenario 12}&$\Cc{9'\mu}^{\rm V}$ & $ +0.2$ & $[+0.07, +0.34]$ &
\multirow{ 2}{*}{1.2} & \multirow{ 2}{*}{3.8\,\%} \\
&$\Cc{10}^{\rm U}$ & $-0.11$ & $[-0.29, +0.07]$ &\\
\hline
\multirow{ 4}{*}{Scenario 13}&$\Cc{9\mu}^{\rm V}$ & $-1.14$ & $[-1.31, -0.96]$ &
\multirow{ 4}{*}{4.7} & \multirow{ 4}{*}{25.4\,\%} \\
&$\Cc{9'\mu}^{\rm V}$ & $  +0.09$ & $[-0.28, +0.45]$ &\\
&$\Cc{10}^{\rm U}$ & $+0.08$ & $[-0.07, +0.24]$ &\\
&$\Cc{10'}^{\rm U}$ & $-0.02$ & $[-0.21, +0.18]$ &\\
\hline\hline
\multirow{ 2}{*}{Scenario 14}&$\Cc{9}^{\rm U}$ & $-1.18$ & $[-1.34, -1.00]$ &
\multirow{ 2}{*}{5.5} & \multirow{ 2}{*}{32.4\,\%} \\
&$\Cc{9'\mu}^{\rm V}$ & $+0.07$ & $[-0.16, +0.30]$ &\\
\hline
\multirow{ 2}{*}{Scenario 15}&$\Cc{9}^{\rm U}$ & $ -1.19$ & $[-1.35, -1.02]$ &
\multirow{ 2}{*}{5.5} & \multirow{ 2}{*}{33.0\,\%} \\
&$\Cc{10'\mu}^{\rm V}$ & $-0.08$ & $[-0.19, +0.04]$ &\\
\hline
\end{tabular}
\end{adjustbox}
\caption{``Electron-less'' fits. Most prominent LFU 2D patterns of NP in $b\to s\ell^+\ell^-$ excluding all electronic observables measured by LHCb.}\label{Fit3Dbis} 
\end{table}

\begin{table}[H] 
    \centering
    \begin{adjustbox}{width=1.\textwidth,center=\textwidth}
\begin{tabular}{c||c|c|c|c|c|c}
 & $\Cc7^{\rm NP}$ & $\Cc{9}^{\rm U}$ & $\Cc{10}^{\rm U}$ & $\Cc{7^\prime}$ & $\Cc{9^\prime}^{\rm U}$ & $\Cc{10^\prime}^{\rm U}$  \\
\hline\hline
Best fit & +0.00 & -1.22 & +0.06 & +0.02 & -0.08 & -0.07 \\ \hline
1$\sigma$ & $[-0.01,+0.02]$ & $[-1.39,-1.04 ]$ & $[-0.09, +0.21]$ & $[+0.00,+0.03]$ & $[-0.44, +0.29]$ &$[-0.26, +0.12]$ 
\end{tabular}
\end{adjustbox}
\caption{``Electron-less'' fits. Most prominent LFU 6D patterns of NP in $b\to s\ell^+\ell^-$ excluding all electronic observables measured by LHCb. The Pull$_{\rm SM}$ is $4.7\sigma$ and the \textit{p}-value is $28.9\%$.}
\label{tab:results6D_noe2}
\end{table}


\begin{table}[H]
 \begin{adjustbox}{width=1.\textwidth,center=\textwidth}
\begin{tabular}{|c|c|c|c|c|}%
\hline
Scenario&Best fit&$1\sigma$&$\text{Pull}_\text{SM}$ &p-value \\%
\hline%
 $(\mathcal{C}_{9\mu}^\text{V}=-\mathcal{C}_{10\mu}^\text{V},\mathcal{C}_{9}^\text{U})$&({-}0.18, {-}0.76)&({[}{-}0.3, {-}0.07{]}, {[}{-}0.88, {-}0.63{]})&6.2&$28.2\%$\\%
\hline%
\end{tabular}
\end{adjustbox}
\caption{``Electron-less'' fits. Outcome of the fit under Scenario 8 including $R_{D^{(*)}}$ and excluding all electronic observables measured by LHCb.}
\label{tab:scenario8impelectronless}
\end{table}

\section{Updated SM predictions}\label{app:SMpredictions}

We provide here the updated SM predictions in the theoretical framework of the full update described in Sec.~\ref{sec:fullupdate}, which includes also
an update of the experimental inputs. We also provide additional predictions for observables that are expected to be measured in the near future.

\begin{longtable}{@{}cccr@{}}
\toprule[1.6pt] 
\multicolumn{4}{c}{Standard Model Predictions} \\
\toprule[1.6pt] 
$10^7\times B(B^+\to K^+\mu\mu)$ [LHCb] & SM & Experiment \cite{LHCb:2014cxe} & Pull\\
 \midrule 
$[0.1, 0.98]$ & $0.32 \pm 0.03$ & $0.29 \pm 0.02$  & $+0.9$ \\
$[1.1, 2.]$ & $0.33 \pm 0.03$ & $0.21 \pm 0.02$  & $+4.0$ \\
$[2., 3.]$ & $0.37 \pm 0.03$ & $0.28 \pm 0.02$  & $+2.5$ \\
$[3., 4.]$ & $0.37 \pm 0.03$ & $0.25 \pm 0.02$  & $+3.4$ \\
$[4., 5.]$ & $0.37 \pm 0.03$ & $0.22 \pm 0.02$  & $+4.4$ \\
$[5., 6.]$ & $0.37 \pm 0.03$ & $0.23 \pm 0.02$  & $+4.0$ \\
$[6., 7.]$ & $0.37 \pm 0.03$ & $0.25 \pm 0.02$  & $+3.3$ \\
$[7., 8.]$ & $0.38 \pm 0.04$ & $0.23 \pm 0.02$  & $+3.1$ \\
$[15., 22.]$ & $1.15 \pm 0.16$ & $0.85 \pm 0.05$  & $+1.8$ \\
\toprule[1.6pt] 
$10^7\times B(B^0\to K^0\mu\mu)$ [LHCb] & SM & Experiment \cite{LHCb:2014cxe} & Pull\\
 \midrule 
$[0.1, 2.]$ & $0.65 \pm 0.05$ & $0.23 \pm 0.11$  & $+3.4$ \\
$[2., 4.]$ & $0.69 \pm 0.05$ & $0.37 \pm 0.11$  & $+2.5$ \\
$[4., 6.]$ & $0.69 \pm 0.05$ & $0.35 \pm 0.11$  & $+2.8$ \\
$[6., 8.]$ & $0.69 \pm 0.07$ & $0.54 \pm 0.12$  & $+1.1$ \\
$[15., 22.]$ & $1.07 \pm 0.15$ & $0.67 \pm 0.12$  & $+2.1$ \\
\toprule[1.6pt] 
$10^7\times B(B^0\to K^{*0}\mu\mu)$ [LHCb] & SM & Experiment \cite{LHCb:2016ykl} & Pull\\
 \midrule 
$[0.1, 0.98]$ & $0.76 \pm 0.45$ & $0.89 \pm 0.09$  & $-0.3$ \\
$[1.1, 2.5]$ & $0.50 \pm 0.25$ & $0.46 \pm 0.06$  & $+0.2$ \\
$[2.5, 4.]$ & $0.53 \pm 0.26$ & $0.50 \pm 0.06$  & $+0.1$ \\
$[4., 6.]$ & $0.81 \pm 0.44$ & $0.71 \pm 0.07$  & $+0.2$ \\
$[6., 8.]$ & $0.97 \pm 0.62$ & $0.86 \pm 0.08$  & $+0.2$ \\
$[15., 19.]$ & $2.54 \pm 0.23$ & $1.74 \pm 0.14$  & $+3.0$ \\
\toprule[1.6pt] 
$10^7\times B(B^+\to K^{*+}\mu\mu)$ [LHCb] & SM & Experiment \cite{LHCb:2014cxe} & Pull\\
 \midrule 
$[0.1, 2.]$ & $1.19 \pm 0.67$ & $1.12 \pm 0.27$  & $+0.1$ \\
$[2., 4.]$ & $0.76 \pm 0.37$ & $1.12 \pm 0.32$  & $-0.7$ \\
$[4., 6.]$ & $0.88 \pm 0.48$ & $0.50 \pm 0.20$  & $+0.7$ \\
$[6., 8.]$ & $1.06 \pm 0.67$ & $0.66 \pm 0.22$  & $+0.6$ \\
$[15., 19.]$ & $2.74 \pm 0.25$ & $1.60 \pm 0.32$  & $+2.8$ \\
\toprule[1.6pt] 
$10^7\times B(B_s\to \phi\mu\mu)$ [LHCb] & SM & Experiment \cite{LHCb:2021zwz} & Pull\\
 \midrule 
$[0.1, 0.98]$ & $1.10 \pm 0.56$ & $0.68 \pm 0.06$  & $+0.7$ \\
$[1.1, 2.5]$ & $0.71 \pm 0.32$ & $0.44 \pm 0.05$  & $+0.9$ \\
$[2.5, 4.]$ & $0.74 \pm 0.37$ & $0.35 \pm 0.04$  & $+1.0$ \\
$[4., 6.]$ & $1.11 \pm 0.70$ & $0.62 \pm 0.06$  & $+0.7$ \\
$[6., 8.]$ & $1.32 \pm 1.09$ & $0.63 \pm 0.06$  & $+0.6$ \\
$[15., 19.]$ & $2.34 \pm 0.15$ & $1.81 \pm 0.12$  & $+2.7$ \\
\toprule[1.6pt] 
$F_L(B^0\to K^{*0}\mu\mu)$ [LHCb] & SM & Experiment \cite{LHCb:2020lmf} & Pull\\
 \midrule 
$[0.1, 0.98]$ & $0.26 \pm 0.09$ & $0.26 \pm 0.03$  & $+0.1$ \\
$[1.1, 2.5]$ & $0.74 \pm 0.07$ & $0.66 \pm 0.05$  & $+0.9$ \\
$[2.5, 4.]$ & $0.80 \pm 0.06$ & $0.76 \pm 0.05$  & $+0.6$ \\
$[4., 6.]$ & $0.73 \pm 0.09$ & $0.68 \pm 0.04$  & $+0.5$ \\
$[6., 8.]$ & $0.65 \pm 0.14$ & $0.65 \pm 0.03$  & $+0.1$ \\
$[15., 19.]$ & $0.34 \pm 0.03$ & $0.35 \pm 0.02$  & $-0.1$ \\
\toprule[1.6pt] 
$P_1(B^0\to K^{*0}\mu\mu)$ [LHCb] & SM & Experiment \cite{LHCb:2020lmf} & Pull\\
 \midrule 
$[0.1, 0.98]$ & $0.05 \pm 0.08$ & $0.09 \pm 0.12$  & $-0.3$ \\
$[1.1, 2.5]$ & $0.03 \pm 0.04$ & $-0.62 \pm 0.30$  & $+2.2$ \\
$[2.5, 4.]$ & $-0.12 \pm 0.09$ & $0.17 \pm 0.37$  & $-0.8$ \\
$[4., 6.]$ & $-0.21 \pm 0.12$ & $0.09 \pm 0.24$  & $-1.1$ \\
$[6., 8.]$ & $-0.24 \pm 0.12$ & $-0.07 \pm 0.21$  & $-0.7$ \\
$[15., 19.]$ & $-0.64 \pm 0.06$ & $-0.58 \pm 0.10$  & $-0.6$ \\
\toprule[1.6pt] 
$P_2(B^0\to K^{*0}\mu\mu)$ [LHCb] & SM & Experiment \cite{LHCb:2020lmf} & Pull\\
 \midrule 
$[0.1, 0.98]$ & $0.13 \pm 0.01$ & $0.00 \pm 0.04$  & $+3.0$ \\
$[1.1, 2.5]$ & $0.45 \pm 0.01$ & $0.44 \pm 0.10$  & $+0.0$ \\
$[2.5, 4.]$ & $0.12 \pm 0.12$ & $0.19 \pm 0.12$  & $-0.4$ \\
$[4., 6.]$ & $-0.26 \pm 0.07$ & $-0.11 \pm 0.07$  & $-1.7$ \\
$[6., 8.]$ & $-0.41 \pm 0.04$ & $-0.21 \pm 0.05$  & $-3.2$ \\
$[15., 19.]$ & $-0.36 \pm 0.02$ & $-0.36 \pm 0.02$  & $-0.1$ \\
\toprule[1.6pt] 
$P_3(B^0\to K^{*0}\mu\mu)$ [LHCb] & SM & Experiment \cite{LHCb:2020lmf} & Pull\\
 \midrule 
$[0.1, 0.98]$ & $-0.00 \pm 0.00$ & $-0.07 \pm 0.06$  & $+1.2$ \\
$[1.1, 2.5]$ & $-0.01 \pm 0.01$ & $-0.32 \pm 0.15$  & $+2.2$ \\
$[2.5, 4.]$ & $-0.01 \pm 0.01$ & $-0.05 \pm 0.20$  & $+0.2$ \\
$[4., 6.]$ & $-0.00 \pm 0.00$ & $0.09 \pm 0.14$  & $-0.7$ \\
$[6., 8.]$ & $-0.00 \pm 0.00$ & $0.07 \pm 0.10$  & $-0.7$ \\
$[15., 19.]$ & $0.00 \pm 0.02$ & $-0.05 \pm 0.05$  & $+1.0$ \\
\toprule[1.6pt] 
$P_4'(B^0\to K^{*0}\mu\mu)$ [LHCb] & SM & Experiment \cite{LHCb:2020lmf} & Pull\\
 \midrule 
$[0.1, 0.98]$ & $-0.46 \pm 0.16$ & $-0.27 \pm 0.24$  & $-0.7$ \\
$[1.1, 2.5]$ & $0.10 \pm 0.14$ & $0.16 \pm 0.29$  & $-0.2$ \\
$[2.5, 4.]$ & $0.77 \pm 0.13$ & $0.87 \pm 0.35$  & $-0.3$ \\
$[4., 6.]$ & $1.02 \pm 0.08$ & $0.62 \pm 0.23$  & $+1.6$ \\
$[6., 8.]$ & $1.09 \pm 0.06$ & $1.15 \pm 0.19$  & $-0.3$ \\
$[15., 19.]$ & $1.28 \pm 0.02$ & $1.28 \pm 0.12$  & $+0.0$ \\
\toprule[1.6pt] 
$P_5'(B^0\to K^{*0}\mu\mu)$ [LHCb] & SM & Experiment \cite{LHCb:2020lmf} & Pull\\
 \midrule 
$[0.1, 0.98]$ & $0.68 \pm 0.14$ & $0.52 \pm 0.10$  & $+0.9$ \\
$[1.1, 2.5]$ & $0.20 \pm 0.13$ & $0.37 \pm 0.12$  & $-0.9$ \\
$[2.5, 4.]$ & $-0.43 \pm 0.12$ & $-0.15 \pm 0.15$  & $-1.5$ \\
$[4., 6.]$ & $-0.72 \pm 0.08$ & $-0.44 \pm 0.12$  & $-1.9$ \\
$[6., 8.]$ & $-0.81 \pm 0.08$ & $-0.58 \pm 0.09$  & $-1.9$ \\
$[15., 19.]$ & $-0.57 \pm 0.05$ & $-0.67 \pm 0.06$  & $+1.2$ \\
\toprule[1.6pt] 
$P_6'(B^0\to K^{*0}\mu\mu)$ [LHCb] & SM & Experiment \cite{LHCb:2020lmf} & Pull\\
 \midrule 
$[0.1, 0.98]$ & $-0.06 \pm 0.02$ & $0.02 \pm 0.09$  & $-0.8$ \\
$[1.1, 2.5]$ & $-0.08 \pm 0.02$ & $-0.23 \pm 0.13$  & $+1.1$ \\
$[2.5, 4.]$ & $-0.06 \pm 0.02$ & $-0.16 \pm 0.15$  & $+0.6$ \\
$[4., 6.]$ & $-0.04 \pm 0.01$ & $-0.29 \pm 0.12$  & $+2.2$ \\
$[6., 8.]$ & $-0.02 \pm 0.01$ & $-0.16 \pm 0.10$  & $+1.4$ \\
$[15., 19.]$ & $-0.00 \pm 0.07$ & $0.07 \pm 0.07$  & $-0.8$ \\
\toprule[1.6pt] 
$P_8'(B^0\to K^{*0}\mu\mu)$ [LHCb] & SM & Experiment \cite{LHCb:2020lmf} & Pull\\
 \midrule 
$[0.1, 0.98]$ & $0.03 \pm 0.02$ & $0.01 \pm 0.24$  & $+0.1$ \\
$[1.1, 2.5]$ & $0.05 \pm 0.03$ & $0.73 \pm 0.32$  & $-2.1$ \\
$[2.5, 4.]$ & $0.05 \pm 0.02$ & $-0.07 \pm 0.34$  & $+0.4$ \\
$[4., 6.]$ & $0.03 \pm 0.01$ & $-0.33 \pm 0.25$  & $+1.4$ \\
$[6., 8.]$ & $0.02 \pm 0.01$ & $0.26 \pm 0.20$  & $-1.2$ \\
$[15., 19.]$ & $-0.00 \pm 0.03$ & $-0.02 \pm 0.14$  & $+0.1$ \\
\toprule[1.6pt] 
$F_L(B^+\to K^{*+}\mu\mu)$ [LHCb] & SM & Experiment \cite{LHCb:2020gog} & Pull\\
 \midrule 
$[0.1, 0.98]$ & $0.27 \pm 0.10$ & $0.34 \pm 0.12$  & $-0.4$ \\
$[1.1, 2.5]$ & $0.75 \pm 0.07$ & $0.54 \pm 0.19$  & $+1.0$ \\
$[2.5, 4.]$ & $0.80 \pm 0.05$ & $0.17 \pm 0.24$  & $+2.5$ \\
$[4., 6.]$ & $0.74 \pm 0.09$ & $0.67 \pm 0.14$  & $+0.4$ \\
$[6., 8.]$ & $0.66 \pm 0.14$ & $0.39 \pm 0.21$  & $+1.0$ \\
$[15., 19.]$ & $0.34 \pm 0.03$ & $0.40 \pm 0.13$  & $-0.4$ \\
\toprule[1.6pt] 
$P_1(B^+\to K^{*+}\mu\mu)$ [LHCb] & SM & Experiment \cite{LHCb:2020gog} & Pull\\
 \midrule 
$[0.1, 0.98]$ & $0.05 \pm 0.08$ & $0.44 \pm 0.41$  & $-0.9$ \\
$[1.1, 2.5]$ & $0.03 \pm 0.04$ & $1.60 \pm 4.93$  & $-0.3$ \\
$[2.5, 4.]$ & $-0.13 \pm 0.10$ & $-0.29 \pm 1.45$  & $+0.1$ \\
$[4., 6.]$ & $-0.21 \pm 0.12$ & $-1.24 \pm 1.21$  & $+0.9$ \\
$[6., 8.]$ & $-0.24 \pm 0.12$ & $-0.78 \pm 0.70$  & $+0.8$ \\
$[15., 19.]$ & $-0.64 \pm 0.06$ & $-0.70 \pm 0.44$  & $+0.1$ \\
\toprule[1.6pt] 
$P_2(B^+\to K^{*+}\mu\mu)$ [LHCb] & SM & Experiment \cite{LHCb:2020gog} & Pull\\
 \midrule 
$[0.1, 0.98]$ & $0.13 \pm 0.02$ & $0.05 \pm 0.12$  & $+0.6$ \\
$[1.1, 2.5]$ & $0.45 \pm 0.01$ & $0.28 \pm 0.45$  & $+0.4$ \\
$[2.5, 4.]$ & $0.10 \pm 0.11$ & $-0.03 \pm 0.28$  & $+0.4$ \\
$[4., 6.]$ & $-0.27 \pm 0.06$ & $0.15 \pm 0.21$  & $-1.9$ \\
$[6., 8.]$ & $-0.41 \pm 0.04$ & $0.06 \pm 0.14$  & $-3.2$ \\
$[15., 19.]$ & $-0.36 \pm 0.02$ & $-0.34 \pm 0.10$  & $-0.2$ \\
\toprule[1.6pt] 
$P_3(B^+\to K^{*+}\mu\mu)$ [LHCb] & SM & Experiment \cite{LHCb:2020gog} & Pull\\
 \midrule 
$[0.1, 0.98]$ & $-0.00 \pm 0.00$ & $0.42 \pm 0.22$  & $-2.0$ \\
$[1.1, 2.5]$ & $-0.01 \pm 0.01$ & $0.09 \pm 1.01$  & $-0.1$ \\
$[2.5, 4.]$ & $-0.01 \pm 0.01$ & $0.45 \pm 0.65$  & $-0.7$ \\
$[4., 6.]$ & $-0.00 \pm 0.00$ & $0.52 \pm 0.83$  & $-0.6$ \\
$[6., 8.]$ & $-0.00 \pm 0.00$ & $-0.17 \pm 0.34$  & $+0.5$ \\
$[15., 19.]$ & $0.00 \pm 0.02$ & $0.07 \pm 0.13$  & $-0.5$ \\
\toprule[1.6pt] 
$P_4'(B^+\to K^{*+}\mu\mu)$ [LHCb] & SM & Experiment \cite{LHCb:2020gog} & Pull\\
 \midrule 
$[0.1, 0.98]$ & $-0.45 \pm 0.16$ & $0.18 \pm 0.76$  & $-0.8$ \\
$[1.1, 2.5]$ & $0.12 \pm 0.15$ & $-1.16 \pm 1.26$  & $+1.0$ \\
$[2.5, 4.]$ & $0.77 \pm 0.13$ & $1.62 \pm 2.20$  & $-0.4$ \\
$[4., 6.]$ & $1.01 \pm 0.08$ & $1.58 \pm 0.96$  & $-0.6$ \\
$[6., 8.]$ & $1.09 \pm 0.06$ & $0.86 \pm 0.91$  & $+0.3$ \\
$[15., 19.]$ & $1.28 \pm 0.02$ & $0.78 \pm 0.47$  & $+1.1$ \\
\toprule[1.6pt] 
$P_5'(B^+\to K^{*+}\mu\mu)$ [LHCb] & SM & Experiment \cite{LHCb:2020gog} & Pull\\
 \midrule 
$[0.1, 0.98]$ & $0.68 \pm 0.14$ & $0.51 \pm 0.32$  & $+0.5$ \\
$[1.1, 2.5]$ & $0.17 \pm 0.12$ & $0.88 \pm 0.72$  & $-1.0$ \\
$[2.5, 4.]$ & $-0.45 \pm 0.11$ & $-0.87 \pm 1.68$  & $+0.2$ \\
$[4., 6.]$ & $-0.73 \pm 0.08$ & $-0.25 \pm 0.41$  & $-1.1$ \\
$[6., 8.]$ & $-0.81 \pm 0.07$ & $-0.15 \pm 0.41$  & $-1.6$ \\
$[15., 19.]$ & $-0.57 \pm 0.05$ & $-0.24 \pm 0.17$  & $-1.9$ \\
\toprule[1.6pt] 
$P_6'(B^+\to K^{*+}\mu\mu)$ [LHCb] & SM & Experiment \cite{LHCb:2020gog} & Pull\\
 \midrule 
$[0.1, 0.98]$ & $-0.06 \pm 0.02$ & $-0.02 \pm 0.40$  & $-0.1$ \\
$[1.1, 2.5]$ & $-0.06 \pm 0.02$ & $0.25 \pm 1.32$  & $-0.2$ \\
$[2.5, 4.]$ & $-0.05 \pm 0.02$ & $-0.37 \pm 3.91$  & $+0.1$ \\
$[4., 6.]$ & $-0.03 \pm 0.01$ & $-0.09 \pm 0.41$  & $+0.1$ \\
$[6., 8.]$ & $-0.02 \pm 0.01$ & $-0.74 \pm 0.40$  & $+1.8$ \\
$[15., 19.]$ & $-0.00 \pm 0.07$ & $-0.28 \pm 0.19$  & $+1.4$ \\
\toprule[1.6pt] 
$P_8'(B^+\to K^{*+}\mu\mu)$ [LHCb] & SM & Experiment \cite{LHCb:2020gog} & Pull\\
 \midrule 
$[0.1, 0.98]$ & $0.09 \pm 0.03$ & $-0.90 \pm 1.02$  & $+1.0$ \\
$[1.1, 2.5]$ & $0.08 \pm 0.03$ & $-0.24 \pm 1.52$  & $+0.2$ \\
$[2.5, 4.]$ & $0.05 \pm 0.02$ & $-0.24 \pm 15.80$  & $+0.0$ \\
$[4., 6.]$ & $0.03 \pm 0.01$ & $0.30 \pm 0.97$  & $-0.3$ \\
$[6., 8.]$ & $0.02 \pm 0.01$ & $0.78 \pm 0.78$  & $-1.0$ \\
$[15., 19.]$ & $-0.00 \pm 0.03$ & $0.22 \pm 0.38$  & $-0.6$ \\
\toprule[1.6pt] 
$P_1(B_s\to \phi\mu\mu)$ [LHCb] & SM & Experiment \cite{LHCb:2021xxq} & Pull\\
 \midrule 
$[0.1, 0.98]$ & $0.11 \pm 0.08$ & $-0.01 \pm 0.19$  & $+0.6$ \\
$[0., 4.]$ & $0.01 \pm 0.06$ & $-0.24 \pm 0.45$  & $+0.5$ \\
$[4., 6.]$ & $-0.19 \pm 0.12$ & $-1.11 \pm 0.50$  & $+1.8$ \\
$[6., 8.]$ & $-0.24 \pm 0.13$ & $0.08 \pm 0.44$  & $-0.7$ \\
$[15., 18.9]$ & $-0.69 \pm 0.03$ & $-0.77 \pm 0.14$  & $+0.6$ \\
\toprule[1.6pt] 
$P_4'(B_s\to \phi\mu\mu)$ [LHCb] & SM & Experiment \cite{LHCb:2021xxq} & Pull\\
 \midrule 
$[0.1, 0.98]$ & $-0.45 \pm 0.15$ & $-0.99 \pm 0.39$  & $+1.3$ \\
$[0., 4.]$ & $0.42 \pm 0.16$ & $0.50 \pm 0.36$  & $-0.2$ \\
$[4., 6.]$ & $1.01 \pm 0.08$ & $0.98 \pm 0.41$  & $+0.1$ \\
$[6., 8.]$ & $1.09 \pm 0.07$ & $0.73 \pm 0.33$  & $+1.1$ \\
$[15., 18.9]$ & $1.30 \pm 0.01$ & $0.87 \pm 0.20$  & $+2.1$ \\
\toprule[1.6pt] 
$P_6'(B_s\to \phi\mu\mu)$ [LHCb] & SM & Experiment \cite{LHCb:2021xxq} & Pull\\
 \midrule 
$[0.1, 0.98]$ & $-0.07 \pm 0.02$ & $-0.41 \pm 0.16$  & $+2.1$ \\
$[0., 4.]$ & $-0.08 \pm 0.02$ & $-0.23 \pm 0.17$  & $+0.9$ \\
$[4., 6.]$ & $-0.04 \pm 0.01$ & $0.39 \pm 0.20$  & $-2.1$ \\
$[6., 8.]$ & $-0.02 \pm 0.01$ & $0.07 \pm 0.17$  & $-0.5$ \\
$[15., 18.9]$ & $-0.00 \pm 0.07$ & $0.01 \pm 0.10$  & $-0.1$ \\
\toprule[1.6pt] 
$F_L(B_s\to \phi\mu\mu)$ [LHCb] & SM & Experiment \cite{LHCb:2021xxq} & Pull\\
 \midrule 
$[0.1, 0.98]$ & $0.26 \pm 0.09$ & $0.25 \pm 0.05$  & $+0.0$ \\
$[0., 4.]$ & $0.77 \pm 0.07$ & $0.72 \pm 0.06$  & $+0.5$ \\
$[4., 6.]$ & $0.74 \pm 0.14$ & $0.70 \pm 0.05$  & $+0.3$ \\
$[6., 8.]$ & $0.66 \pm 0.26$ & $0.62 \pm 0.05$  & $+0.1$ \\
$[15., 18.9]$ & $0.35 \pm 0.02$ & $0.36 \pm 0.04$  & $-0.1$ \\
\toprule[1.6pt] 
$F_L(B^0\to K^{*0}ee)$ [LHCb] & SM & Experiment \cite{LHCb:2020dof} & Pull\\
 \midrule 
$[0.0008, 0.257]$ & $0.04 \pm 0.02$ & $0.04 \pm 0.03$  & $-0.2$ \\
\toprule[1.6pt] 
$P_1(B^0\to K^{*0}ee)$ [LHCb] & SM & Experiment \cite{LHCb:2020dof} & Pull\\
 \midrule 
$[0.0008, 0.257]$ & $0.04 \pm 0.09$ & $0.11 \pm 0.10$  & $-0.5$ \\
\toprule[1.6pt] 
$P_2(B^0\to K^{*0}ee)$ [LHCb] & SM & Experiment \cite{LHCb:2020dof} & Pull\\
 \midrule 
$[0.0008, 0.257]$ & $0.01 \pm 0.00$ & $0.03 \pm 0.04$  & $-0.5$ \\
\toprule[1.6pt] 
$P_3(B^0\to K^{*0}ee)$ [LHCb] & SM & Experiment \cite{LHCb:2020dof} & Pull\\
 \midrule 
$[0.0008, 0.257]$ & $-0.00 \pm 0.00$ & $0.01 \pm 0.05$  & $-0.2$ \\
\toprule[1.6pt] 
$R_K$ [LHCb] & SM & Experiment \cite{LHCb:2022zom} & Pull\\
 \midrule 
$[0.1, 1.1]$ & $0.99 \pm 0.00$ & $0.99 \pm 0.09$  & $-0.0$ \\
$[1.1, 6.]$ & $1.00 \pm 0.00$ & $0.95 \pm 0.05$  & $+1.1$ \\
\toprule[1.6pt] 
$R_{K_S}$ [LHCb] & SM & Experiment \cite{LHCb:2021lvy} & Pull\\
 \midrule 
$[1.1, 6.]$ & $1.00 \pm 0.00$ & $0.66 \pm 0.20$  & $+1.7$ \\
\toprule[1.6pt] 
$R_K$ [Belle] & SM & Experiment \cite{BELLE:2019xld} & Pull\\
 \midrule 
$[1., 6.]$ & $1.00 \pm 0.00$ & $1.03 \pm 0.28$  & $-0.1$ \\
$[14.18, 22.9]$ & $1.00 \pm 0.00$ & $1.16 \pm 0.30$  & $-0.5$ \\
\toprule[1.6pt] 
$R_{K^*}$ [LHCb] & SM & Experiment \cite{LHCb:2022zom} & Pull\\
 \midrule 
$[0.1, 1.1]$ & $0.98 \pm 0.00$ & $0.93 \pm 0.10$  & $+0.5$ \\
$[1.1, 6.]$ & $1.00 \pm 0.00$ & $1.03 \pm 0.08$  & $-0.4$ \\
\toprule[1.6pt] 
$R_{K^{*+}}$ [LHCb] & SM & Experiment \cite{LHCb:2021lvy} & Pull\\
 \midrule 
$[0.045, 6.]$ & $0.94 \pm 0.02$ & $0.70 \pm 0.18$  & $+1.3$ \\
\toprule[1.6pt] 
$R_{K^*}$ [Belle] & SM & Experiment \cite{Belle:2019oag} & Pull\\
 \midrule 
$[0.045, 1.1]$ & $0.91 \pm 0.01$ & $0.52 \pm 0.36$  & $+1.1$ \\
$[1.1, 6.]$ & $1.00 \pm 0.00$ & $0.96 \pm 0.46$  & $+0.1$ \\
$[15., 19.]$ & $1.00 \pm 0.00$ & $1.18 \pm 0.53$  & $-0.3$ \\
\toprule[1.6pt] 
$P_4'(B\to K^{*}ee)$ [Belle] & SM & Experiment \cite{Belle:2016fev} & Pull\\
 \midrule 
$[0.1, 4.]$ & $0.02 \pm 0.14$ & $-0.68 \pm 0.93$  & $+0.7$ \\
$[4., 8.]$ & $1.05 \pm 0.07$ & $1.04 \pm 0.48$  & $+0.0$ \\
$[14.18, 19.]$ & $1.27 \pm 0.03$ & $0.30 \pm 0.82$  & $+1.2$ \\
\toprule[1.6pt] 
$P_4'(B\to K^{*}\mu\mu)$ [Belle] & SM & Experiment \cite{Belle:2016fev} & Pull\\
 \midrule 
$[0.1, 4.]$ & $0.06 \pm 0.14$ & $0.76 \pm 1.03$  & $-0.7$ \\
$[4., 8.]$ & $1.05 \pm 0.07$ & $0.14 \pm 0.66$  & $+1.4$ \\
$[14.18, 19.]$ & $1.27 \pm 0.03$ & $0.20 \pm 0.79$  & $+1.3$ \\
\toprule[1.6pt] 
$P_5'(B\to K^{*}ee)$ [Belle] & SM & Experiment \cite{Belle:2016fev} & Pull\\
 \midrule 
$[0.1, 4.]$ & $0.20 \pm 0.10$ & $0.51 \pm 0.47$  & $-0.6$ \\
$[4., 8.]$ & $-0.77 \pm 0.08$ & $-0.52 \pm 0.28$  & $-0.8$ \\
$[14.18, 19.]$ & $-0.60 \pm 0.05$ & $-0.91 \pm 0.36$  & $+0.8$ \\
\toprule[1.6pt] 
$P_5'(B\to K^{*}\mu\mu)$ [Belle] & SM & Experiment \cite{Belle:2016fev} & Pull\\
 \midrule 
$[0.1, 4.]$ & $0.18 \pm 0.10$ & $0.42 \pm 0.41$  & $-0.6$ \\
$[4., 8.]$ & $-0.77 \pm 0.08$ & $-0.03 \pm 0.32$  & $-2.2$ \\
$[14.18, 19.]$ & $-0.60 \pm 0.05$ & $-0.13 \pm 0.39$  & $-1.2$ \\
\toprule[1.6pt] 
$Q_4(B\to K^{*})$ [Belle] & SM & Experiment \cite{Belle:2016fev} & Pull\\
 \midrule 
$[0.1, 4.]$ & $0.05 \pm 0.01$ & $1.46 \pm 1.39$  & $-1.0$ \\
$[4., 8.]$ & $0.00 \pm 0.00$ & $-0.90 \pm 0.80$  & $+1.1$ \\
$[14.18, 19.]$ & $0.00 \pm 0$ & $-0.08 \pm 1.14$  & $+0.1$ \\
\toprule[1.6pt] 
$Q_5(B\to K^{*})$ [Belle] & SM & Experiment \cite{Belle:2016fev} & Pull\\
 \midrule 
$[0.1, 4.]$ & $-0.01 \pm 0.01$ & $-0.10 \pm 0.62$  & $+0.1$ \\
$[4., 8.]$ & $-0.00 \pm 0.00$ & $0.50 \pm 0.42$  & $-1.2$ \\
$[14.18, 19.]$ & $0 \pm 0$ & $0.78 \pm 0.51$  & $-1.5$ \\
\toprule[1.6pt] 
$10^7\times B(B^+\to K^+\mu\mu)$ [Belle] & SM & Experiment \cite{BELLE:2019xld} & Pull\\
 \midrule 
$[1., 6.]$ & $1.85 \pm 0.14$ & $2.30 \pm 0.41$  & $-1.0$ \\
$[14.18, 22.9]$ & $1.38 \pm 0.20$ & $1.34 \pm 0.24$  & $+0.1$ \\
\toprule[1.6pt] 
$10^7\times B(B^0\to K^0\mu\mu)$ [Belle] & SM & Experiment \cite{BELLE:2019xld} & Pull\\
 \midrule 
$[1., 6.]$ & $1.72 \pm 0.13$ & $0.62 \pm 0.44$  & $+2.4$ \\
$[14.18, 22.9]$ & $1.29 \pm 0.18$ & $0.98 \pm 0.44$  & $+0.6$ \\
\toprule[1.6pt] 
$F_L(B^0\to K^{*0}\mu\mu)$ [ATLAS] & SM & Experiment \cite{ATLAS:2018gqc} & Pull\\
 \midrule 
$[0.04, 2.]$ & $0.36 \pm 0.10$ & $0.44 \pm 0.11$  & $-0.5$ \\
$[2., 4.]$ & $0.80 \pm 0.05$ & $0.64 \pm 0.12$  & $+1.2$ \\
$[4., 6.]$ & $0.73 \pm 0.09$ & $0.42 \pm 0.18$  & $+1.6$ \\
\toprule[1.6pt] 
$P_1(B^0\to K^{*0}\mu\mu)$ [ATLAS] & SM & Experiment \cite{ATLAS:2018gqc} & Pull\\
 \midrule 
$[0.04, 2.]$ & $0.05 \pm 0.08$ & $-0.05 \pm 0.31$  & $+0.3$ \\
$[2., 4.]$ & $-0.10 \pm 0.08$ & $-0.78 \pm 0.61$  & $+1.1$ \\
$[4., 6.]$ & $-0.21 \pm 0.12$ & $0.14 \pm 0.50$  & $-0.7$ \\
\toprule[1.6pt] 
$P_4'(B^0\to K^{*0}\mu\mu)$ [ATLAS] & SM & Experiment \cite{ATLAS:2018gqc} & Pull\\
 \midrule 
$[0.04, 2.]$ & $-0.29 \pm 0.14$ & $-0.62 \pm 0.89$  & $+0.4$ \\
$[2., 4.]$ & $0.67 \pm 0.14$ & $1.52 \pm 0.75$  & $-1.1$ \\
$[4., 6.]$ & $1.02 \pm 0.08$ & $-1.28 \pm 0.75$  & $+3.0$ \\
\toprule[1.6pt] 
$P_5'(B^0\to K^{*0}\mu\mu)$ [ATLAS] & SM & Experiment \cite{ATLAS:2018gqc} & Pull\\
 \midrule 
$[0.04, 2.]$ & $0.52 \pm 0.11$ & $0.67 \pm 0.31$  & $-0.5$ \\
$[2., 4.]$ & $-0.34 \pm 0.13$ & $-0.33 \pm 0.34$  & $-0.0$ \\
$[4., 6.]$ & $-0.72 \pm 0.08$ & $0.26 \pm 0.39$  & $-2.4$ \\
\toprule[1.6pt] 
$P_6'(B^0\to K^{*0}\mu\mu)$ [ATLAS] & SM & Experiment \cite{ATLAS:2018gqc} & Pull\\
 \midrule 
$[0.04, 2.]$ & $-0.06 \pm 0.02$ & $-0.18 \pm 0.21$  & $+0.5$ \\
$[2., 4.]$ & $-0.07 \pm 0.02$ & $0.31 \pm 0.34$  & $-1.1$ \\
$[4., 6.]$ & $-0.04 \pm 0.01$ & $0.06 \pm 0.30$  & $-0.3$ \\
\toprule[1.6pt] 
$P_8'(B^0\to K^{*0}\mu\mu)$ [ATLAS] & SM & Experiment \cite{ATLAS:2018gqc} & Pull\\
 \midrule 
$[0.04, 2.]$ & $0.03 \pm 0.02$ & $0.58 \pm 1.03$  & $-0.5$ \\
$[2., 4.]$ & $0.05 \pm 0.02$ & $-2.14 \pm 1.13$  & $+1.9$ \\
$[4., 6.]$ & $0.03 \pm 0.01$ & $0.48 \pm 0.86$  & $-0.5$ \\
\toprule[1.6pt] 
$P_1(B^0\to K^{*0}\mu\mu)$ [CMS8] & SM & Experiment \cite{CMS:2017rzx} & Pull\\
 \midrule 
$[1., 2.]$ & $0.05 \pm 0.05$ & $0.12 \pm 0.48$  & $-0.1$ \\
$[2., 4.3]$ & $-0.11 \pm 0.09$ & $-0.69 \pm 0.62$  & $+0.9$ \\
$[4.3, 6]$ & $-0.21 \pm 0.12$ & $0.53 \pm 0.38$  & $-1.9$ \\
$[6, 8.68]$ & $-0.25 \pm 0.12$ & $-0.47 \pm 0.31$  & $+0.7$ \\
$[16., 19.]$ & $-0.70 \pm 0.05$ & $-0.53 \pm 0.25$  & $-0.7$ \\
\toprule[1.6pt] 
$P_5'(B^0\to K^{*0}\mu\mu)$ [CMS8] & SM & Experiment \cite{CMS:2017rzx} & Pull\\
 \midrule 
$[1., 2.]$ & $0.34 \pm 0.12$ & $0.10 \pm 0.33$  & $+0.7$ \\
$[2., 4.3]$ & $-0.38 \pm 0.13$ & $-0.57 \pm 0.38$  & $+0.5$ \\
$[4.3, 6]$ & $-0.73 \pm 0.08$ & $-0.96 \pm 0.33$  & $+0.7$ \\
$[6, 8.68]$ & $-0.82 \pm 0.08$ & $-0.64 \pm 0.23$  & $-0.7$ \\
$[16., 19.]$ & $-0.53 \pm 0.04$ & $-0.56 \pm 0.14$  & $+0.2$ \\
\toprule[1.6pt] 
$F_L(B^0\to K^{*0}\mu\mu)$ [CMS8] & SM & Experiment \cite{CMS:2015bcy} & Pull\\
 \midrule 
$[1., 2.]$ & $0.70 \pm 0.08$ & $0.64 \pm 0.12$  & $+0.4$ \\
$[2., 4.3]$ & $0.79 \pm 0.06$ & $0.80 \pm 0.10$  & $-0.1$ \\
$[4.3, 6]$ & $0.73 \pm 0.09$ & $0.62 \pm 0.12$  & $+0.7$ \\
$[6, 8.68]$ & $0.64 \pm 0.16$ & $0.50 \pm 0.08$  & $+0.8$ \\
$[16., 19.]$ & $0.34 \pm 0.03$ & $0.38 \pm 0.07$  & $-0.6$ \\
\toprule[1.6pt] 
$A_{FB}(B^0\to K^{*0}\mu\mu)$ [CMS8] & SM & Experiment \cite{CMS:2015bcy} & Pull\\
 \midrule 
$[1., 2.]$ & $-0.17 \pm 0.05$ & $-0.27 \pm 0.41$  & $+0.2$ \\
$[2., 4.3]$ & $-0.04 \pm 0.03$ & $-0.12 \pm 0.18$  & $+0.4$ \\
$[4.3, 6.]$ & $0.11 \pm 0.05$ & $0.01 \pm 0.15$  & $+0.6$ \\
$[6., 8.68]$ & $0.22 \pm 0.10$ & $0.03 \pm 0.10$  & $+1.3$ \\
$[16., 19.]$ & $0.34 \pm 0.03$ & $0.35 \pm 0.07$  & $-0.1$ \\
\toprule[1.6pt] 
$10^7\times B(B^0\to K^{*0}\mu\mu)$ [CMS8] & SM & Experiment \cite{CMS:2015bcy} & Pull\\
 \midrule 
$[1., 2.]$ & $0.37 \pm 0.19$ & $0.46 \pm 0.08$  & $-0.4$ \\
$[2., 4.3]$ & $0.81 \pm 0.40$ & $0.76 \pm 0.12$  & $+0.1$ \\
$[4.3, 6.]$ & $0.70 \pm 0.38$ & $0.58 \pm 0.10$  & $+0.3$ \\
$[6., 8.68]$ & $1.35 \pm 0.88$ & $1.26 \pm 0.13$  & $+0.1$ \\
$[16., 19.]$ & $1.76 \pm 0.15$ & $1.26 \pm 0.13$  & $+2.5$ \\
\toprule[1.6pt] 
$F_{H}(B^+\to K^+\mu\mu)$ [CMS8] & SM & Experiment \cite{CMS:2018qih} & Pull\\
 \midrule 
$[1., 2.]$ & $0.05 \pm 0.00$ & $0.21 \pm 0.49$  & $-0.3$ \\
$[2., 4.3]$ & $0.02 \pm 0.00$ & $0.85 \pm 0.37$  & $-2.3$ \\
$[4.3, 8.68]$ & $0.01 \pm 0.00$ & $0.01 \pm 0.04$  & $+0.0$ \\
$[16., 18.]$ & $0.01 \pm 0.00$ & $0.07 \pm 0.10$  & $-0.6$ \\
$[18., 22.]$ & $0.01 \pm 0.00$ & $0.10 \pm 0.13$  & $-0.7$ \\
\toprule[1.6pt] 
$A_{FB}(B^+\to K^+\mu\mu)$ [CMS8] & SM & Experiment \cite{CMS:2018qih} & Pull\\
 \midrule 
$[1., 2.]$ & $0 \pm 0$ & $0.08 \pm 0.23$  & $-0.4$ \\
$[2., 4.3]$ & $0 \pm 0$ & $-0.04 \pm 0.14$  & $+0.3$ \\
$[4.3, 8.68]$ & $0 \pm 0$ & $0 \pm 0.04$  & $+0$ \\
$[16., 18.]$ & $0 \pm 0$ & $0.04 \pm 0.06$  & $-0.7$ \\
$[18., 22.]$ & $0 \pm 0$ & $0.05 \pm 0.05$  & $-0.9$ \\
\toprule[1.6pt] 
$F_L(B^+\to K^{*+}\mu\mu)$ [CMS8] & SM & Experiment \cite{CMS:2020oqb} & Pull\\
 \midrule 
$[1., 8.68]$ & $0.71 \pm 0.10$ & $0.60 \pm 0.34$  & $+0.3$ \\
$[14.18, 19.]$ & $0.35 \pm 0.04$ & $0.55 \pm 0.14$  & $-1.4$ \\
\toprule[1.6pt] 
$A_{FB}(B^+\to K^{*+}\mu\mu)$ [CMS8] & SM & Experiment \cite{CMS:2020oqb} & Pull\\
 \midrule 
$[1., 8.68]$ & $0.09 \pm 0.04$ & $-0.14 \pm 0.39$  & $+0.6$ \\
$[14.18, 19.]$ & $0.37 \pm 0.03$ & $0.33 \pm 0.12$  & $+0.3$ \\
\toprule[1.6pt] 
$F_L(B^0\to K^{*0}\mu\mu)$ [CMS7] & SM & Experiment \cite{CMS:2013mkz} & Pull\\
 \midrule 
$[1., 2.]$ & $0.70 \pm 0.08$ & $0.60 \pm 0.34$  & $+0.3$ \\
$[2., 4.3]$ & $0.79 \pm 0.06$ & $0.65 \pm 0.17$  & $+0.8$ \\
$[4.3, 8.68]$ & $0.67 \pm 0.13$ & $0.81 \pm 0.14$  & $-0.7$ \\
$[16., 19.]$ & $0.34 \pm 0.03$ & $0.44 \pm 0.08$  & $-1.3$ \\
\toprule[1.6pt] 
$A_{FB}(B^0\to K^{*0}\mu\mu)$ [CMS7] & SM & Experiment \cite{CMS:2013mkz} & Pull\\
 \midrule 
$[1., 2.]$ & $-0.17 \pm 0.05$ & $-0.29 \pm 0.41$  & $+0.3$ \\
$[2., 4.3]$ & $-0.04 \pm 0.03$ & $-0.07 \pm 0.20$  & $+0.1$ \\
$[4.3, 8.68]$ & $0.18 \pm 0.08$ & $-0.01 \pm 0.11$  & $+1.4$ \\
$[16., 19.]$ & $0.34 \pm 0.03$ & $0.41 \pm 0.06$  & $-1.1$ \\
\toprule[1.6pt] 
$10^7\times B(B^0\to K^{*0}\mu\mu)$ [CMS7] & SM & Experiment \cite{CMS:2013mkz} & Pull\\
 \midrule 
$[1., 2.]$ & $0.37 \pm 0.19$ & $0.48 \pm 0.15$  & $-0.4$ \\
$[2., 4.3]$ & $0.81 \pm 0.40$ & $0.87 \pm 0.18$  & $-0.2$ \\
$[4.3, 8.68]$ & $2.05 \pm 1.26$ & $1.62 \pm 0.35$  & $+0.3$ \\
$[16., 19.]$ & $1.76 \pm 0.15$ & $1.56 \pm 0.23$  & $+0.7$ \\
\toprule[1.6pt] 
$10^5\times B(B^0\to K^{*0}\gamma)$ [PDG] & SM & Experiment \cite{ParticleDataGroup:2020ssz} & Pull\\
 \midrule 
 & $3.64 \pm 2.43$ & $4.18 \pm 0.25$ & $-0.2$ \\
\toprule[1.6pt] 
$10^5\times B(B^+\to K^{*+}\gamma)$ [PDG] & SM & Experiment \cite{ParticleDataGroup:2020ssz} & Pull\\
 \midrule 
 & $3.63 \pm 2.51$ & $3.92 \pm 0.22$ & $-0.1$ \\
\toprule[1.6pt] 
$10^5\times B(B_s\to \phi\gamma)$ [PDG] & SM & Experiment \cite{ParticleDataGroup:2020ssz} & Pull\\
 \midrule 
 & $5.26 \pm 3.08$ & $3.40 \pm 0.40$ & $+0.6$ \\
\toprule[1.6pt] 
$ 10^4 \times B(B\to X_s\gamma)\text{[HFLAV]}^\dag $ & SM \cite{Misiak:2020vlo} & Experiment \cite{HeavyFlavorAveragingGroup:2022wzx} & Pull \\ 
 \midrule 
 & $ 3.32 \pm 0.15 $ & $ 3.40 \pm 0.17 $ & $ -0.4 $ \\ 
 \toprule[1.6pt] 
 
$  S(B \to K^* \gamma)\text{[BaBar+Belle]}^\dag $ & SM \cite{Descotes-Genon:2011nqe} & Experiment \cite{HeavyFlavorAveragingGroup:2022wzx} & Pull \\ 
 \midrule 
 & $-0.03\pm 0.01$ & $-0.16 \pm 0.22$ & ${+0.6}$  \\ 
 \toprule[1.6pt] 
$  AI(B \to K^* \gamma)\text{[BaBar+Belle]}^\dag $ & SM \cite{Descotes-Genon:2011nqe} & Experiment \cite{HeavyFlavorAveragingGroup:2022wzx} & Pull \\ 
 \midrule 
 & $0.041\pm0.025$  & $0.063 \pm 0.017$  & ${-0.7}$ \\ 

 \toprule[1.6pt] 
 
$ 10^9\times B(B_s \to \mu^+\mu^-) \text{[LHCb+CMS+ATLAS]}^\dag $ & SM \cite{MisiakOrsay} & Experiment \cite{Hurth:2022lnw} & Pull \\ 
 \midrule 
 & $3.64\pm0.14$ & $3.52\pm0.32$  & $+0.3$  \\ 

 \toprule[1.6pt] 
$ 10^{6} \times B(B\to X_s\mu^+\mu^-)\text{[BaBar]}^\dag $ & SM \cite{Huber:2020vup} & Experiment \cite{BaBar:2013qry} & Pull \\ 
 \midrule 
  $[1,6]$ & $1.73 \pm 0.13$  & $0.66\pm 0.88$ & $+1.2$  \\ 
 \toprule[1.6pt] 
$ 10^{6} \times B(B\to X_s e^+e^-)\text{[BaBar]}^\dag $ & SM \cite{Huber:2020vup} & Experiment \cite{BaBar:2013qry} & Pull \\ 
 \midrule 
  $[1,6]$ & $1.78 \pm 0.13$  & $1.93\pm 0.55$ & $-0.3$  \\ 
\bottomrule[1.6pt]
\end{longtable}

\begin{longtable}{@{}cccc@{}}
\toprule[1.6pt]
\multicolumn{4}{c}{Additional SM Predictions}\\
\midrule
Observables & $[1.1,6.0]$ & $[1.1,7.0]$\\
\midrule
$F_L(B^0\to K^{*0}e^+e^-)$ & $+0.765 \pm 0.070$ & $+0.747 \pm 0.079$ \\
$P_1(B^0\to K^{*0}e^+e^-)$ & $-0.118 \pm 0.089$ & $-0.147 \pm 0.096$ \\
$P_2(B^0\to K^{*0}e^+e^-)$ & $+0.021 \pm 0.092$ & $-0.082 \pm 0.087$ \\
$P_3(B^0\to K^{*0}e^+e^-)$ & $-0.005 \pm 0.004$ & $-0.005 \pm 0.004$ \\
$P^\prime_4(B^0\to K^{*0}e^+e^-)$ & $+0.69 \pm 0.13$ & $+0.77 \pm 0.12$ \\
$P^\prime_5(B^0\to K^{*0}e^+e^-)$ & $-0.38 \pm 0.11$ & $-0.47 \pm 0.11$ \\
$P^\prime_6(B^0\to K^{*0}e^+e^-)$ & $-0.054 \pm 0.017$ & $-0.047 \pm 0.015$ \\
$P^\prime_8(B^0\to K^{*0}e^+e^-)$ & $+0.041 \pm 0.019$ & $+0.037 \pm 0.016$ \\
$Q_{F_L}(B^0\to K^{*0})\times 10^3$ & $-13.8 \pm 2.1$ & $-12.0 \pm 1.9$ \\
$Q_1(B^0\to K^{*0})\times 10^3$ & $-1.47 \pm 0.75$ & $-1.43 \pm 0.69$ \\
$Q_2(B^0\to K^{*0})\times 10^3$ & $-2.36 \pm 0.92$ & $-3.19 \pm 0.52$ \\
$Q_3(B^0\to K^{*0})\times 10^3$ & $+0.003 \pm 0.029$ & $+0.007 \pm 0.026$ \\
$Q_4(B^0\to K^{*0})\times 10^3$ & $+5.55 \pm 0.83$ & $+5.33 \pm 0.94$ \\
$Q_5(B^0\to K^{*0})\times 10^3$ & $-6.60 \pm 0.71$ & $-6.76 \pm 0.64$ \\
$Q_6(B^0\to K^{*0})\times 10^3$ & $-0.28 \pm 0.12$ & $-0.160 \pm 0.082$ \\
$Q_8(B^0\to K^{*0})\times 10^3$ & $-0.078 \pm 0.075$ & $-0.104 \pm 0.082$ \\
\bottomrule[1.6pt]
\end{longtable}
\newpage


\bibliographystyle{JHEP}

\bibliography{reference}

\end{document}